\documentclass[10pt, twocolumn]{article}
\usepackage[margin=0.75in]{geometry}
\usepackage{helvet}
\usepackage{braket}
\usepackage{amsmath}
\usepackage{amssymb}
\usepackage{dsfont}
\usepackage{titlesec}
\titleformat{\section}{\normalfont\sffamily\large\bfseries}{\thesection}{1em}{}
\titleformat{\subsection}{\normalfont\sffamily\bfseries}{\thesubsection}{1em}{}
\usepackage{scicite}
\usepackage{url}
\usepackage[normalem]{ulem}
\usepackage{graphicx}
\renewenvironment{abstract}
	{\quotation}
	{\endquotation}
\date{}

\makeatletter
\renewcommand{\fnum@figure}{\textbf{Figure \thefigure}}
\renewcommand{\fnum@table}{\textbf{Table \thetable}}
\makeatother

\newcommand{\id}{\mathds{1}}

\def\scititle{
	Quantum algorithms for equational reasoning
}
\title{\bfseries \boldmath{ \scititle}}

\author{
	Davide~Rattacaso$^{1,2\ast}$,
	Daniel~Jaschke$^{1,2,3,4}$,
	Marco~Ballarin$^{1,2,5}$,\and
	Ilaria~Siloi$^{1,2}$,
	Simone~Montangero$^{1,2}$\and
	\small$^{1}$Dipartimento di Fisica e Astronomia “G. Galilei” \& Padua Quantum Technologies Research Center,\and \small Università degli Studi di Padova, Italy I-35131, Padova, Italy.\and
	\small$^{2}$INFN, Sezione di Padova, via Marzolo 8, I-35131, Padova.\and
	\small$^{3}$Institute for Complex Quantum Systems, Ulm University, Albert-Einstein-Allee 11, 89069 Ulm, Germany.\and
	\small$^{4}$Current affiliation: PlanQC GmbH, Lichtenbergstr. 8, 85748 Garching, Germany.\and
	\small$^{5}$Current affiliation: Quantinuum, Partnership House, Carlisle Place, London SW1P 1BX, United Kingdom.\and
	\small$^\ast$Corresponding author. Email: davide.rattacaso@unipd.it
}

\begin{document} 

\maketitle

\begin{abstract} \bfseries \boldmath
As a cornerstone of automated reasoning, equational reasoning finds equivalences between symbolic expressions and fuels advances across scientific disciplines. Yet, its potential remains limited by the exponential growth of equivalent expressions with increasing problem size. We introduce quantum normal form reduction, a quantum computational framework designed to address this challenge. We construct an efficiently implementable quantum Hamiltonian whose ground state encodes all equivalent expressions in a quantum superposition. By preparing and manipulating these states, we tackle fundamental problems in equational reasoning, including verifying and counting equivalent expressions and identifying structural properties of equivalence classes. We demonstrate a quantum-inspired version of the algorithm, using tensor networks to solve instances involving up to $10^{28}$ equivalent expressions, far beyond the reach of classical graph exploration. This framework opens the path for quantum symbolic computation in areas from circuit design to data compression, computational group theory, linguistics, and macromolecular modeling, unlocking previously inaccessible problems.
\end{abstract}

\section*{Introduction}

Equivalence relations group individual objects into categories based on shared structure or behavior, enabling reasoning at the level of entire classes rather than isolated instances. This abstraction is central across disciplines: biologists study species rather than individual organisms, mathematicians analyze functions instead of specific representations, physicists examine macrostates rather than microstates, and linguists consider languages beyond single sentences. In symbolic computation, equivalence relations can be encoded and manipulated algorithmically, allowing computers to automate such reasoning. A key approach is provided by term rewriting systems~\cite{10.5555/559395,baader1998term, plaisted1993equational}: given one expression and a set of rewriting rules, all other equivalent expressions are generated by applying a sequence of rule-based substitutions.

\begin{figure*}[t]
\includegraphics[width=\linewidth]{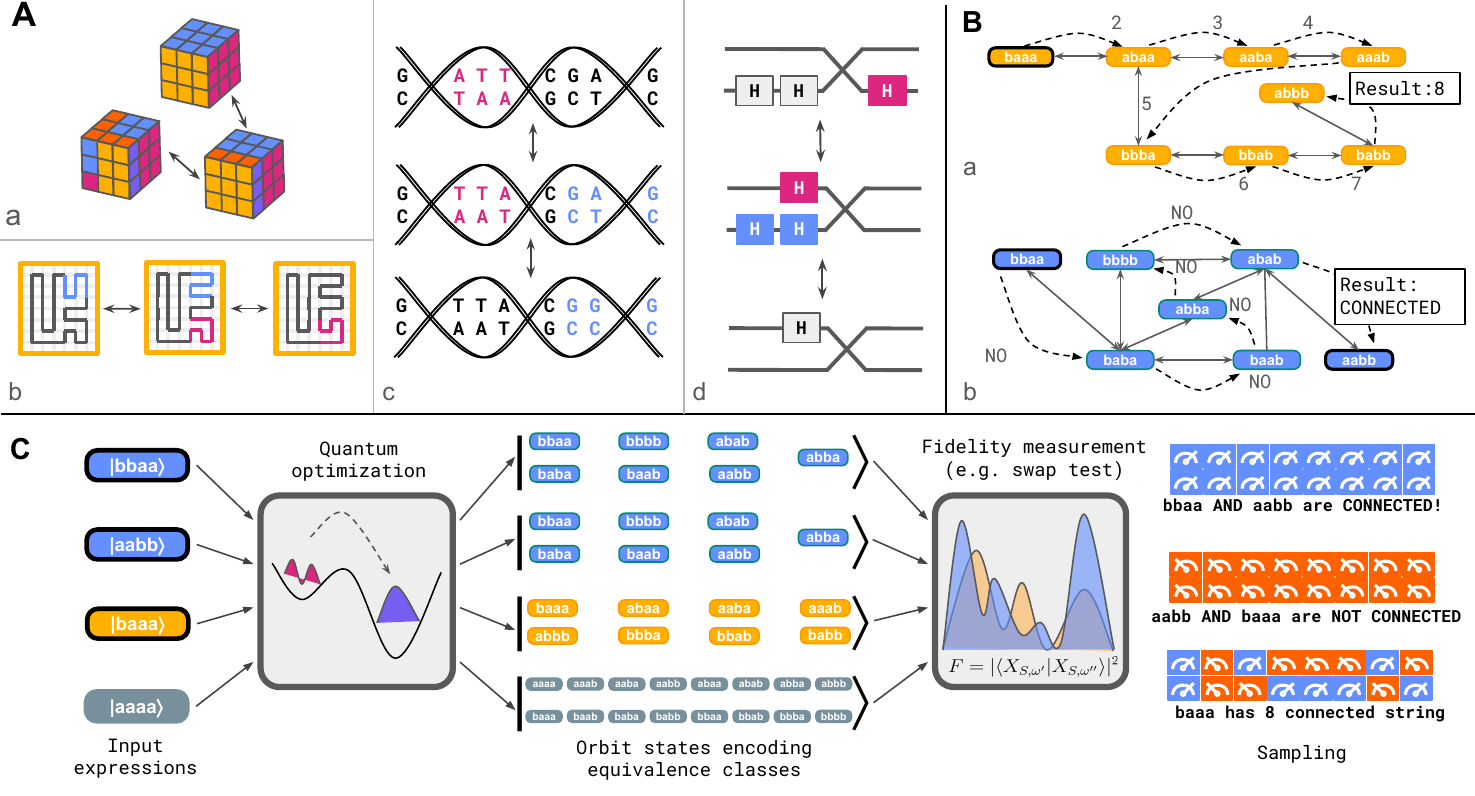}
\caption{\textbf{Overview.} \emph{Panel~A:} Term rewriting enables the exploration of large sets of symbolic data structures that represent objects sharing common properties by connecting them through transformation rules. For example, (a) elements of the same group of transformations, such as the Rubik group, are connected via the relations among the group generators (b) space-filling curves on a lattice are connected via local transformations that preserve topology, (c) genomes of individuals within the same species are connected through a set of genetic mutations, and (d) quantum circuits implementing the same unitary operator are connected via the replacement of equivalent subcircuits.
\emph{Panel~B:} Certain computational tasks over these large sets can be performed classically by sequentially exploring the set through repeated applications of the rewriting rules. The double arrows connect pairs of words that are related by the application of a rule, while the dashed lines represent a specific sequence of words explored by the classical algorithm, to count the number of connected words (a) or determine if some target word is reachable (b). 
\emph{Panel~C:} In quantum normal form reduction, we leverage quantum optimization to reduce an \textit{input expression} into a corresponding \textit{orbit states}, which encode the equivalence class into a quantum state by using equally weighted superpositions of the class members. Global properties of equivalence classes can be extracted efficiently via quantum operations on orbit states. For example, the fidelity between orbit states encodes solutions to the word problem and counting, and can be estimated by sampling the system state with a swap test. Expressions in the same equivalence class share the same color (green, grey, or yellow)}
\label{fig:main}
\end{figure*} 
Across scientific domains, it is well known that a few simple rewriting rules can encapsulate complex semantics and generate rich behaviors. In mathematics, term rewriting provides a unifying framework for encoding algebraic identities, logical equivalences, and inference rules. This underpins symbolic algorithms for solving equations~\cite{baader1998term}, performing computations in algebraic groups and monoids~\cite{Sims_1994} (Figure~\ref{fig:main}Aa) , and proving theorems~\cite{hsiang1985refutational}. In classical and quantum information processing, rewriting rules enable local substitutions of equivalent subcircuits (Figure~\ref{fig:main}Ad)~\cite{10175801}, which is crucial for verifying functional correctness and optimizing resource usage through circuit compilation~\cite{5010420,burgholzer2020advanced, 9384317, Duncan_2020}. Rewriting also extends to formal grammars in the field of linguistics, which define the generative structure of natural language, and in computer science provide the syntactic foundation of programming languages, supporting tasks such as equivalence checking and compiler optimization~\cite{chomsky2002syntactic, 10.5555/1196416,DragonBook}. In data compression, formal grammar-based encoding allows lossless compression of large datasets~\cite{grammar_based_codes}. In automata theory, rewriting systems enable the systematic exploration of configuration spaces, being able to simulate any Turing machine~\cite{10.5555/1196416,baader1998term}. In biology and chemistry, rewriting frameworks capture both the structural and informational aspects of macromolecules such as DNA (Figure~\ref{fig:main}Ac), RNA, proteins, and polymers(Figure~\ref{fig:main}Ab)~\cite{Searls2002,searls2013primer, stoyan1996enumeration}.

The expressive power of rewriting systems is accompanied by intrinsic computational challenges. Iteratively applying rewriting rules to a single object generates an equivalence class that can grow combinatorially. Many foundational and applied scientific investigations in all previous domains require exploring such equivalence classes in full or in part. For example, verifying whether two circuits implement the same function reduces to determining whether they can be transformed into one another using a prescribed set of functionality-preserving rewrites~\cite{10175801}. In computational group theory, this corresponds to the \textit{word problem} for finitely generated groups~\cite{Bone1958}: determining whether two sequences of generators produce the same group element. Related decision problems arise across disciplines. The grammar equivalence problem~\cite{10.5555/1196416, 10.1007/3-540-63165-8_221} --- determining whether two formal grammars generate the same language --- is central to both computational linguistics and programming language theory. In automata theory and statistical physics, a crucial question is to determine how many distinct states a given rewriting system or automaton can explore, or how many space-filling curves are admissible on a given lattice~\cite{bodroza2013enumeration, stoyan1996enumeration}. This counting problem connects to entropy, complexity, and information content, with relevance in fields such as polymer thermodynamics~\cite{Vanderzande_1998}, automata-based models of physical systems~\cite{VICHNIAC198496}, and genetic diversity analysis~\cite{Ellegren2016}.

Due to the ability of rewriting systems to simulate arbitrary computations, the word problem is undecidable in general~\cite{Bone1958}. Only a narrow subset of the aforementioned questions can currently be answered using existing algorithms with feasible resource requirements, also for decidable problems, due to the exponential size of the equivalence classes. But a different form of computation, such as quantum computing, can offer a new theoretical perspective, or in some cases a computational advantage, in the vast field of equational reasoning. Quantum computation is a fundamentally new computational paradigm, enabling polynomial and exponential speedups for problems that are intractable for classical algorithms~\cite{nielsen_chuang, Ladd2010,Montanaro2016}.

Here, we extend quantum computation to automated symbolic reasoning by introducing \textit{quantum normal form reduction}, a quantum computational framework tailored to address a variety of the aforementioned challenges. The key insight is that quantum mechanics enables the representation and manipulation of entire equivalence classes encoded as coherent quantum superpositions of the class members (Figure~\ref{fig:main}C), bypassing the need to sequentially represent individual elements (Figure~\ref{fig:main}B). We refer to these class-representing quantum states as \textit{orbit states}. We demonstrate that these states can be prepared on a quantum computer, and that key questions can be efficiently answered by performing quantum operations on orbit states. For example, the \textit{word problem} is solved by measuring overlap between two orbit states, and the number of elements of an equivalence class is inferred by measuring an appropriate observable. 

As a quantum analogue to normal form reduction (see Supplementary Text for details), which is usually performed by exploiting the Knuth-Bendix algorithm~\cite{Knuth1983}, the preparation of orbit states via quantum normal form reduction allows one to associate a unique quantum state to an entire class of equivalent expressions. We demonstrate that orbit states can be prepared as ground states of an appropriate sparse Hamiltonian, specifically the discrete Laplacian of the configuration graph generated by the action of the rewriting system~\cite{Chung:1997}. This Hamiltonian is constructed as a linear combination of tensor products of local operators, each encoding a rule of the rewriting system. As a consequence, we show that the computational cost of simulating its action on a quantum device scales polynomially with the number of rules in the rewriting system itself and with the size of the rules, i.e., the number of characters on which each rule acts.

Orbit state preparation can be accomplished using current quantum optimization techniques, such as quantum annealing~\cite{Kadowaki_PRE1998,Farhi_SCI01,Santoro_SCI02,aqc_review}, optimal control~\cite{opt_c_0, opt_c_3, opt_c_4}, quantum approximate optimization algorithms (QAOA)~\cite{farhi2014quantum,farhi2019quantum, BLEKOS20241} and imaginary time evolution~\cite{mcardle2019variational, Motta2020, PRXQuantum.3.010320}. Despite the exponential size of the equivalence classes, the amount of quantum memory required to represent the orbit state is polynomial, while the amount of time needed to prepare the state depends on the specific instance of the equivalence problem~\cite{Abbas2024}. Once the state has been prepared, many global properties of the equivalence class can be measured efficiently. Solving the word problem and counting elements inside an equivalence class reduces to measuring the quantum fidelity of orbit states. This operation can be performed in polynomial time, for instance, through a swap test~\cite{PhysRevLett.87.167902}, laying the groundwork for tackling previously inaccessible problems.

We employ Tensor Network (TN) methods~\cite{PhysRevB.94.165116, RevModPhys.77.259, SCHOLLWOCK201196, Montangero2018, Silvi2019} to emulate the execution of the proposed quantum algorithm and to demonstrate its effectiveness. Tensor network methods enable efficient classical simulations of quantum processes, under specific conditions regarding the structure of quantum correlations and entanglement~\cite{Evenbly2011}. The TN implementation effectively defines a quantum-inspired classical algorithm that can be executed on existing classical hardware, already enabling efficient classical analysis, in some cases, beyond the current state of the art.

We focus on a toy term rewriting system in which an initial string of size $L$ is updated by substituting substrings of equal length. The string length $L$ corresponds to the quantum memory requirements of our device. We address both the word problem and the counting problem for input strings of different sizes. Despite the exponential size of the equivalence classes, we efficiently represent and manipulate orbit states generated from strings of length up to $L = 100$, containing up to $10^{28}$ connected strings encoded in $1$ gigabyte of memory. The memory required to encode the same data as a list of strings would be approximately $ 10^{17}$ terabytes. This demonstrates the potential of our algorithm as a powerful compression method for sets of data that are related by rewriting relations.

\section*{Results}

\subsection*{Term rewriting systems}

A term rewriting system consists of a set of rewrite rules that operate on syntactic terms, capturing how complex structures evolve through rule-based replacements. Term rewriting can manipulate a variety of symbolic data structures, including strings, lattice configurations, and graphs. Since any data structure can be serialized and ultimately represented as a bit string, we focus on rewriting strings. As a consequence, rewriting rules that act locally on the original data structure might assume a non-local representation when applied to the serialized data structure. In this setting, we denote as a string rewriting system a term rewriting system in which all rewriting rules act locally on strings.

Thus, without loss of generality, we define an invertible term rewriting system $S $ as  
\begin{equation}
    S = \left[A\Big| R\right]\;,
\end{equation}  
where
\begin{equation}
    A = \{\alpha_i, i \in [1,\dots,d]\}
\end{equation}
is an alphabet of $d$ characters used to compose strings, $A^*$ is the set of all the possible strings over the alphabet $A$, and
\begin{equation}
    R = \{ r_l\; |\; r_l:A^*\rightarrow A^*,\; l \in [1,\dots,n_r]\}
\end{equation}  
is a set of $n_r$ rewriting rules, i.e., functions mapping a string being in the set $A^*$ to another string in $A^*$ via replacement operations on a set of characters. In particular, a rule  
\begin{equation}
    r_l = (\alpha_{[j_1]} \beta_{[j_2]}\dots \approx  \alpha'_{[j_1]} \beta'_{[j_2]}\dots)
\end{equation}
simultaneously replaces the character $\alpha $ at position $j_1 $ with $\alpha' $, the character $\beta $ at position $j_2 $ with $\beta' $, and so on. The equivalence symbol $\approx$ indicates that we restrict to invertible rules, meaning that the reverse transformation can also be applied. Here, we consider rules that preserve the length of strings. More general rules can be made length-preserving by the introduction of a \textit{blank} character.

Invertible rules establish an equivalence relation among connected strings, then partitioning the space of strings into disjoint equivalence classes~\cite{baader1998term}. Each equivalence class consists of all strings that are mutually reachable via arbitrary sequences of rewriting rules from the set $S$.  To label these classes, we choose a representative element $\tilde{\omega}$ within a given class and denote by $X_{S,\tilde{\omega}}$ the set of all strings that can be obtained by applying sequences of rules from $S$ to the initial string $\tilde{\omega}$. As an illustrative example, consider the rewriting system
\begin{equation}
S = \left[\{a,b\} \mid \{a_1b_2\approx b_1a_2, a_2b_3\approx b_2a_3\}\right],
\end{equation}
acting on strings of fixed length $L=3$. The space of all length-$3$ strings is partitioned by the rules of $S$ into four equivalence classes:
$X_{S,aaa}=\{aaa\}$,
$X_{S,aab}=\{aab, aba, baa\}$,
$X_{S,abb}=\{abb, bab, bba\}$,
and
$X_{S,bbb}=\{bbb\}$, where we labeled each equivalence class by choosing the lexicographically smallest element as its representative. For this rewriting system, the word problem, namely, determining whether two given words are equivalent, has a positive answer for the pair $aab$ and $baa$, and a negative answer for the pair $aab$ and $aaa$. The counting problem, which asks for the number of words equivalent to a given input, yields $1$ for $aaa$ and $bbb$, and $3$ for $aab$ and $bab$.

More elaborate decision problems arise when comparing multiple rewriting systems. For instance, consider a second system
\begin{equation}
S' = \left[\{a,b\} \mid \{a_1a_2\approx b_1b_2,, a_2a_3\approx b_2b_3\}\right],
\end{equation}
and ask whether $S$ and $S'$ generate the same equivalence class when applied to the same input word $aaa$. This question, which captures a simple instance of the grammar equivalence problem~\cite{10.5555/1196416, 10.1007/3-540-63165-8_221}, has a negative answer in this case, since
$X_{S',aaa}=\{aaa, abb, bba\}\neq X_{S,aaa}$. 

The rewriting systems in these examples exhibit particularly simple word and counting problems because their dynamics preserve an easily identifiable invariant. In the first system, this invariant is the total number of $a$ characters, while in the second it is the parity of that number. In contrast, equivalence relations relevant to more complex settings, such as equivalence problems for circuits, cannot generally be reduced to the conservation of a simple quantity. This increased complexity is largely due to the presence of context-dependent rules, that is, rules that apply only when a specific local pattern is matched. Within our formalism, the rule $aaa \approx aba$ constitutes an example of a context-dependent rewriting rule, since the central character $a$ is replaced only when surrounded by other characters $a$.

\subsection*{Quantum states and orbit states}

Quantum computing is the use of controllable quantum many-body systems to process information and solve computational problems. We consider many-body systems composed of $L$ local subsystems, each associated with a finite set of $d$ discrete classical configurations, such as the two orientations of a spin in a magnetic field. When local measurements of all subsystems are performed on such a system, the readout yields a single classical configuration $\omega_k$ with $k\in[1,\dots,d^L]$, assigning to each local subsystem one of its classical configurations. Repeating the same quantum computation multiple times produces a statistical ensemble of outcomes, effectively sampling configurations $\omega_k$ with associated probabilities $p_k$. The goal of quantum computation is to manipulate the quantum system so that the solution to a problem of interest can be efficiently extracted from the measurement statistics.

What distinguishes quantum computation from classical probabilistic computation is the nature of the system’s evolution prior to measurement. While stochastic processes evolve probability distributions with non-negative real weights, quantum systems are described by complex probability amplitudes that may interfere constructively or destructively. In particular, the state of the system is represented by a normalized vector $\ket{\psi}$ in a Hilbert space constructed by associating a computational basis vector $\ket{\omega_k}$ to each classical configuration $\omega_k$,
\begin{equation}
\ket{\psi} = \sum_k \psi_k \ket{\omega_k},
\end{equation}
where the $\psi_k$ are complex amplitudes. The probability $p_k$ of observing the configuration $\omega_k$ upon measurement is given by the Born rule, $p_k = |\psi_k|^2$ and the measurement irreversibly projects the state onto the observed configuration, eliminating all other components of the superposition.

For implementing symbolic quantum computation, any string of $L$ characters from the alphabet $A$ is mapped to a configuration of a many-body quantum system. To this aim, we define a quantum system composed of $L$ local subsystems with $d$ internal states, or \textit{qudits}, where $d = |A|$ is the size of the alphabet. Each internal state of the qudit is labeled by a character in $A$. Upon measuring such a quantum system in the computational basis, we sample a classical configuration $(\alpha_{k_1},\dots,\alpha_{k_L})$ encoding a string $\omega_k = \alpha_{k_1}\dots\alpha_{k_L}$, and we assert that the system is in the state $\ket{\omega_k}=\ket{\alpha_{k_1},\dots,\alpha_{k_L}}$. As we will show hereafter, before measurement, the state of the system is in general in a quantum superposition of states, specifically $\ket{\psi}=\sum_k \psi_k\ket{\omega_k}$. The Hilbert space has dimension $d^L$, corresponding to the maximum number of different strings that can be sampled.

Each rewriting rule $r_l\in R$ naturally defines a linear operator $\hat{r}_l$ that acts on a quantum superposition of strings by applying the rule $r$ to each computational basis state. For a rule $r_l=(\alpha_{[j_1]} \beta_{[j_2]}\dots \approx  \alpha'_{[j_1]} \beta'_{[j_2]}\dots)$, the corresponding operator can be written in the bra-ket notation as:
\begin{equation}\label{eq:rule_operator}
    \hat{r}_l = \ket{\alpha'}\bra{\alpha}_{j_1}\otimes\ket{\beta'}\bra{\beta}_{j_2}\otimes\cdots +  \ket{\alpha}\bra{\alpha'}_{j_1}\otimes\ket{\beta}\bra{\beta'}_{j_2}\otimes\cdots,
\end{equation}
where $\ket{\alpha}\bra{\alpha'}_j$ is the local operator that replace the character $\alpha'$ with the character $\alpha$ in any computational basis state containing the character $\alpha'$ at position $j$, while multiplies by $0$ other computational basis states. The two terms in the latter expression correspond to the two directions in which the rule can be applied, and together ensure that the operator is Hermitian, as required for a valid quantum observable.

As shown in the previous section, the rewriting rules in the term rewriting system $S$ partition the space of all strings of a given size into disjoint equivalence classes. We encode each equivalence class $X_{S,\tilde{\omega}}$ as an equally weighted quantum superposition of all words $\omega \in X_{S,\tilde{\omega}}$, which we designate as the orbit state:  
\begin{equation}\label{eq:orbit_state_def}
    \ket{X_{S,\tilde{\omega}}} = \sum_{\omega\in X_{S,\tilde{\omega}}}\frac{1}{\sqrt{|X_{S,\tilde{\omega}}|}}\ket{\omega}\;.
\end{equation}
As we will show, once the orbit state has been prepared, one can efficiently manipulate the entire equivalence class via quantum operations.

\subsection*{A parent Hamiltonian for orbit states}

The dynamics of a closed quantum system is governed by a Hermitian operator acting on its state, known as the Hamiltonian. The Hamiltonian encodes the energetic structure of the system: its eigenvalues correspond to the possible energy levels, and its eigenvectors to the quantum states with definite energy. One important paradigm of quantum computation is to encode the solution of a computational problem into the lowest-energy eigenstate, or ground state, of a suitably designed Hamiltonian. The desired quantum state can then be prepared using physical processes such as adiabatic evolution.

\begin{figure}[t]
\includegraphics[width=\linewidth]{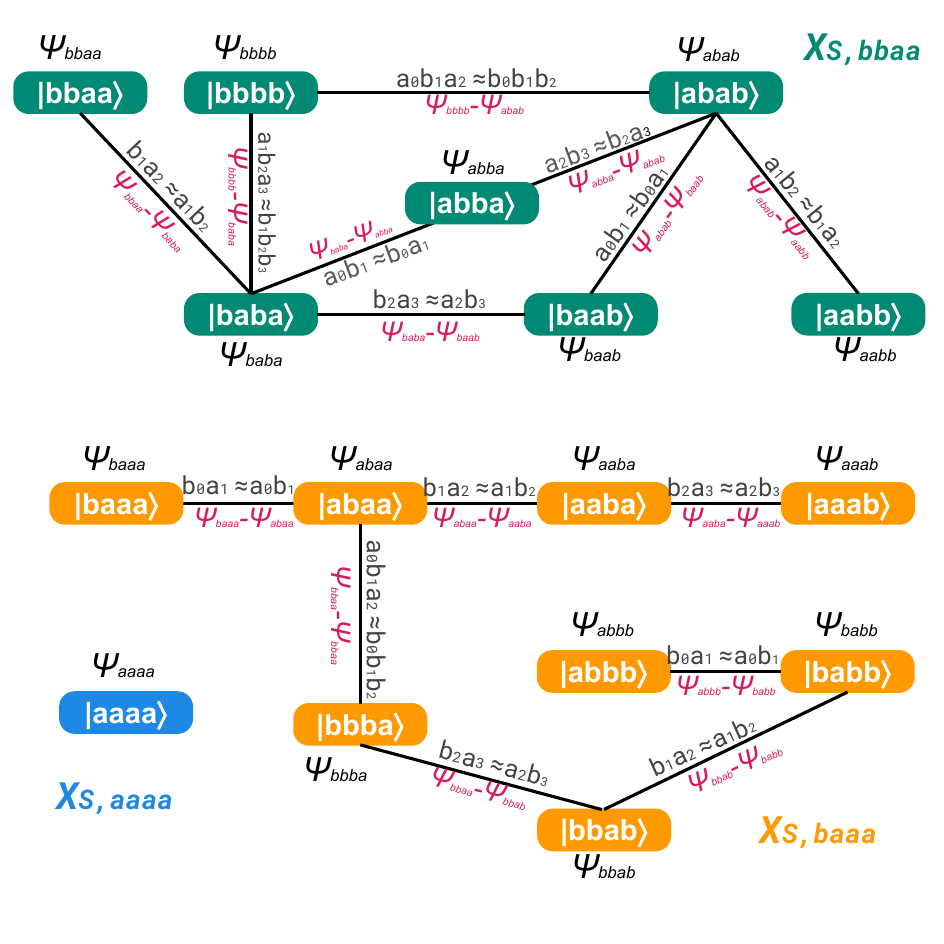}
\caption{\textbf{Discrete calculus over the disconnected graph defined by the action of a rewriting system on strings of size $L=4$.} The rewriting system is $S=\big[\{a,b\},\{a_ib_{i+1}\approx b_ia_{i+1}, a_ib_{i+1}a_{i+2}\approx b_ib_{i+1}b_{i+2}\}\big]$. The action of the rewriting rules on strings defines a graph, in which two strings are connected by an edge if one can be obtained from the other via the application of a rule. This graph consists of three connected subgraphs shown in green, yellow, and blue, whose vertices correspond to the three equivalence classes $X_{S,\text{bbaa}}$, $X_{S,\text{baaa}}$, and $X_{S,\text{aaaa}}$. A complex function $\psi_k$ defined on the vertices of the graph specifies a quantum state $\ket{\psi} = \sum_k \psi_k \ket{\omega_k}$. The discrete gradient $\vec{\nabla}\psi$ quantifies the variation of $\psi$ along each edge $(\omega_k, \omega_{k'})$, given by the difference $\psi_{k'} - \psi_k$ in red along the edges.}
\label{fig:graph}
\end{figure}

We construct a parent Hamiltonian $\mathcal{L}_{S}$ whose degenerate ground states are precisely the orbit states defined in the previous section. While finding a parent Hamiltonian is generally a challenging task~\cite{PhysRevLett.132.160401,PhysRevLett.122.150606,
Fernández-González2015}, in this case, we systematically build $\mathcal{L}_{S}$ directly from the rewriting rules in $S$, exploiting the notion of discrete Laplacian on a graph~\cite{Chung:1997} (see Figure~\ref{fig:graph}).

We define the graph $G=(V,E)$, where the vertices $V = \{\omega_k\}$ represent the possible strings of size $L$. The edges $E=\{(\omega_k,\omega_{k'})\}$ connect strings that are related by a rewriting rule $r$. By construction, the graph $G$ is a disconnected graph with the equivalence classes forming connected subgraphs. Note that the size of the graph $G$ increases exponentially with the string length $L$. Since each vertex represents a string and each string corresponds to a state of the computational basis, a quantum state $\ket{\psi}=\sum_k \psi_k\ket{\omega_k}$ can be represented as complex function $\psi: V \rightarrow\mathbb{C}$ that associates to each vertex $\omega_k$ of the graph the complex value $\psi_k$.

Orbit states correspond to the flattest complex functions on the graph $G$, as they remain constant in both modulus and phase on each connected component. For a function defined on a graph, flatness can be quantified as the minimization of differences between values assigned to adjacent vertices.  The variation of $\psi$ along edges is measured by its discrete gradient $\vec{\nabla} \psi$ (in red in Figure~\ref{fig:graph}), that is the function that associates to each edge $(\omega_k,\omega_{k'})\in E$ the difference $\psi_{k'}-\psi_{k}$.
The sum of $|\psi_{k'}-\psi_{k}|^2$ on all edges measures the total variation of $\psi$, namely its Dirichlet energy~\cite{Chung:1997}. It reads:
\begin{align}\label{eq:dirichelet}
E_\text{D} &= \sum_{(\omega_k,\omega_{k'})\in E}|\psi_{k}-\psi_{k'}|^2 \nonumber\\
&= \sum_{(\omega_k,\omega_{k'})\in E}(|\psi_{k}|^2-\psi_{k'}^*\psi_{k} -\psi_{k'}\psi_{k}^*  + |\psi_{k'}|^2)\;.
\end{align}

In the Supplementary Text, we show that the Dirichlet energy associated with a function $\psi$ is given by the expectation value of a positive semi-definite graph Laplacian operator $\hat{\mathcal{L}}_S$ on the state $\ket{\psi}$:
\begin{equation}
    E_\text{D} = \langle \psi|\hat{\mathcal{L}}_S|\psi\rangle\;.
\end{equation}
In this framework, orbit states are ground states of the Laplacian and have zero energy. The operator $\hat{\mathcal{L}}_S$ is sparse and defined by the rewriting rules $r_l \in R$ of the system $S$:
\begin{equation}
    \hat{\mathcal{L}}_S= \sum_{\hat{r}_l\;:\;r_l\in R}\left( \hat{r_l}^2 - \hat{r}_l\right)\;,
\end{equation}
where $\hat{r}_l$ is the operator encoding the rule $r_l$, as defined in Eq.~(\ref{eq:rule_operator}).

Although both the size of the graph $G$ and the dimension of the Hilbert space grow exponentially with the string length $L$, the action of the Laplacian operator $\hat{\mathcal{L}}_S$ can nevertheless be simulated efficiently on a quantum computer. More precisely, for a term rewriting system specified by a number of rules that scales at most polynomially with $L$, the time evolution generated by $\hat{\mathcal{L}}_S$ can be approximated to fixed accuracy by a quantum circuit of depth scaling polynomially in $L$. This efficiency follows from the fact that the Laplacian decomposes into a sum of rule operators, each of which acts as a product of single-qudit operators (see Supplementary Text for details).

\subsection*{Preparing the orbit state}

While the ground state space of $\hat{\mathcal{L}}_S$ is spanned by arbitrary superpositions of orbit states, our goal is to prepare a specific orbit state $\ket{X_{S,\tilde{\omega}}}$. A natural strategy is to drive the system toward the ground state of the Laplacian while constraining the dynamics to the subspace of the Hilbert space spanned solely by computational-basis states corresponding to strings reachable from $\tilde{\omega}$. To achieve this, we add to the Laplacian the projector $P_{\tilde{\omega}} = -\ket{\tilde{\omega}}\bra{\tilde{\omega}}$. This addition leads to the family of Hamiltonians:  
\begin{equation}
    \hat{H}_{S,\tilde{\omega}}(h) = (1-h)\hat{\mathcal{L}}_S - h \ket{\tilde{\omega}}\bra{\tilde{\omega}}.
\end{equation}
The off-diagonal component of this family of Hamiltonians is composed of the rule operators of the rewriting system. As a consequence, the dynamics generated by $\hat{H}_{S,\tilde{\omega}}(h)$ acting on the initial state $\ket{\tilde{\omega}}$ is confined to the subspace spanned by the equivalent strings in $X_{S,\tilde{\omega}}$. Within this subspace, the unique ground state is the orbit state $\ket{X_{S,\tilde{\omega}}}$.

To prepare the orbit state on a quantum computer or simulator, one possible strategy is to design an evolution that follows the instantaneous ground states $\ket{\phi(h)}$ of the Hamiltonian $\hat{H}_{S,\tilde{\omega}}(h)$ as $h$ transitions from 1 to 0. If the evolution is adiabatic, the final state at the end of this process is then
\begin{equation}
\lim_{h\rightarrow0}\ket{\phi(h)}=\ket{X_{S,\tilde{\omega}}}.
\end{equation}

As well as the Laplacian operator, the Hamiltonian $\hat{H}_{S,\tilde{\omega}}$ can be simulated with a quantum circuit involving a number of gates that scales polynomially with the number of rules in $S$ and their size (see Supplementary Text). Thus, the evolution can be implemented using state-of-the-art quantum algorithms such as Quantum Annealing (QA)~\cite{Kadowaki_PRE1998,Farhi_SCI01,Santoro_SCI02,aqc_review}, Optimal Control~\cite{opt_c_0, opt_c_3, opt_c_4}, Quantum Approximate Optimization Algorithms~\cite{farhi2014quantum,farhi2019quantum,BLEKOS20241} or Imaginary Time Evolution (ITE)~\cite{mcardle2019variational, Motta2020, PRXQuantum.3.010320} and shortcuts to adiabaticity~\cite{del_Campo_2013, Gu_ry_Odelin_2019}.

In a QA-based approach, the initial state $\ket{\psi(t=0)}=\ket{\tilde{\omega}}$ evolves under the time-dependent Hamiltonian
\begin{equation}
    \hat H(t) = \hat{H}_{S,\tilde{\omega}}\left(\frac{\tau-t}{\tau}\right) = \frac{t}{\tau}\hat{\mathcal{L}}_S - \left(\frac{\tau-t}{\tau}\right) \ket{\tilde{\omega}}\bra{\tilde{\omega}}
\end{equation}
for $t:0\rightarrow\tau$. If $\tau$ is sufficiently large, the evolution adiabatically follows the ground state path
\begin{equation}
    \ket{\psi(t)} = \ket{\phi\left(\frac{\tau-t}{\tau}\right)}\;,
\end{equation}
whose final state is the target orbit state.

Here, to benchmark the proposed approach, we perform a TN simulation of the algorithm, focusing on a mixture of QA and ITE, which we refer to as Imaginary Quantum Annealing (IQA)~\cite{PhysRevB.93.224431}.  In an IQA-based approach, the evolution is governed by the imaginary Hamiltonian
\begin{equation}
    \hat I(t) = i\,\hat{H}_{S,\tilde{\omega}}\left(\frac{\tau-t}{\tau}\right)\;,
\end{equation}
and the system follows the adiabatic path by gradually suppressing the amplitude of the excited components generated by non-perfect adiabatic evolution. A key advantage of IQA is that, at any given time, the imaginary Hamiltonian dampens previously generated excitations, effectively mitigating errors and eventually outperforming standard QA~\cite{PhysRevB.93.224431}. In a tensor network (TN) simulation, damping can also suppress excitations that arise from approximating intermediate highly entangled states with finite bond dimension.

\subsection*{Quantum algorithms for equational reasoning}

Many global properties of equivalence classes can be efficiently extracted via quantum operations on orbit states.

We begin by focusing on the reconstruction of the overlap function
\begin{equation}
    F(X_{S_1,\omega_1}, X_{S_2,\omega_2}) = \frac{|X_{S_1,\omega_1} \cap X_{S_2,\omega_2}|^2}{|X_{S_1,\omega_1}| \cdot |X_{S_2,\omega_2}|},
\end{equation}
where $X_{S_1,\omega_1}$ is the equivalence class generated by the action of a term rewriting system $S_1$ on the input word $\omega_1$, $X_{S_2,\omega_2}$ is the equivalence class generated by the action of a possibly different term rewriting system $S_2$ on the input word $\omega_2$, and $|X|$ is the number of elements in $X$. The function $F(X_{S_1,\omega_1}, X_{S_2,\omega_2})$ quantifies the squared size of the intersection between the two equivalence classes $X_{S_1,\omega_1}$ and $X_{S_2,\omega_2}$, normalized by the product of their sizes. This quantity measures the similarity between equivalence classes: it approaches zero when the overlap is negligible and reaches one when the two sets coincide. As we show here, the knowledge of $F$ enables the resolution of several important problems, such as the word problem, the counting problem, and the grammar-equivalence problem.

\subsubsection*{Measuring equivalence classes overlap via fidelity}

Once two orbit states $\ket{X_{S_1,\omega_1}}$ and $\ket{X_{S_2,\omega_2}}$ are prepared in two memory registers of a digital quantum computer as shown in the previous section, their similarity can be quantified by measuring the squared magnitude of their overlap,  i.e., $\left|\langle{X_{S_1,\omega_1}}\ket{X_{S_2,\omega_2}}\right|^2$. This function is also called \textit{quantum fidelity}~\cite{nielsen_chuang}. Using the definition in Eq.~(\ref{eq:orbit_state_def}), the fidelity between orbit states $\ket{X_{S_1,\omega_1}}$ and $\ket{X_{S_2,\omega_2}}$ is reduced to the overlap function $F(X_{S_1,\omega_1}, X_{S_2,\omega_2})$ as follows:
\begin{align}\label{eq:orbit_fidelity}
    &|\langle X_{S_1,\omega_1} \ket{X_{S_2,\omega_2}}|^2 \nonumber \\
      &= \frac{1}{|X_{S_1,\omega_1}|\cdot|X_{S_2,\omega_2}|} \left| \sum_{\substack{\omega \in X_{S_1,\omega_1} \\ \omega' \in X_{S_2,\omega_2}}} \langle \omega \ket{\omega'} \right|^2 \nonumber \\
      &= \frac{|X_{S_1,\omega_1} \cap X_{S_2,\omega_2}|^2}{|X_{S_1,\omega_1}| \cdot |X_{S_2,\omega_2}|}\nonumber\\
      &=F\;,
\end{align}
since only the non-zero terms in the summation correspond to the strings that simultaneously belong to both equivalence classes. The ratio $F(X_{S_1,\omega_1}, X_{S_2,\omega_2})$ can be efficiently measured using the swap test algorithm~\cite{PhysRevLett.87.167902}, which requires only a polynomial number of gates relative to the length of the input words. Furthermore, as discussed in Supplementary Text, fidelity between orbit states can also be efficiently estimated on analog quantum simulators such as quantum annealers.

\subsubsection*{Word problem}
When $S_1 = S_2$, the strings $\omega_1$ and $\omega_2$ belong to the same equivalence class if and only if they are connected under the rewriting system. In this case,
\begin{equation}
    F = 
    \begin{cases}
        1 & \text{if } \omega_1 \text{ and } \omega_2 \text{ are connected,} \\
        0 & \text{otherwise,}
    \end{cases}
\end{equation}
thus providing a solution to the word problem.

\subsubsection*{Counting problem}
Let us consider the rewriting system
\begin{equation}
    S_A = \left[A,R=\{\alpha_i\approx\alpha'_i\;|\;\alpha,\alpha'\in A, i\in[1,\dots,L]\}\right]\;,
\end{equation}
which replaces an arbitrary character at any position with any other character. $S_A$ is thus capable of generating all the possible strings over the alphabet. Consequently, we can construct a uniform superposition over all such strings in the Hilbert space as an orbit state for $S_A$, i.e.,
\begin{equation}\label{eq:counting_fraction}
    \ket{All} = \ket{X_{S_A,\omega_1}} = \frac{1}{\sqrt{d^L}}\sum_{\omega} \ket{\omega},
\end{equation}
where $\omega_1$ is an arbitrary input string. 
In this special case, the function $F=|\langle X_{S_A,\omega_1} \ket{X_{S_2,\omega_2}}|^2$ reduces to
\begin{equation}
    F = \frac{|X_{S_2,\omega_2}|}{d^L},
\end{equation}
which allows us to estimate the number of strings connected to $\omega_2$ under the action of $S_2$, thus solving the counting problem. Importantly, the uniform superposition state $\ket{All}$ can be efficiently implemented on digital quantum computers using Hadamard gates as commonly done in most quantum algorithms~\cite{nielsen_chuang}.

\subsubsection*{Filtering}

Another relevant operation on equivalence classes is the extraction of a specific subset of elements that satisfy a given condition. We name this operation filtering. For instance, one might be interested in generating all space-filling curves on a square lattice that exhibit inversion symmetry. The ability to filter elements is also essential for applications in formal grammars, where it may be necessary to exclude strings containing nonterminal symbols.
\\We define the target subset as
\begin{equation}
    X_{S,\tilde{\omega},g} = \{\omega \in X_{S,\tilde{\omega}} \mid g(\omega) = 1\},
\end{equation}
where \( g \) is a Boolean function that returns 1 if and only if the string \( \omega \) satisfies the desired property. Once the orbit state has been prepared, subset extraction can be implemented by introducing an ancilla qubit initialized in the state \( \ket{0} \), which is flipped conditionally based on the value of \( g(\omega) \). This results in the following transformation:
\begin{equation}
    \sum_{\omega\in X_{S,\tilde{\omega}}}\frac{1}{\sqrt{|X_{S,\tilde{\omega}}|}}\ket{\omega}\otimes\ket{0}
    \longrightarrow
    \sum_{\omega\in X_{S,\tilde{\omega}}}\frac{1}{\sqrt{|X_{S,\tilde{\omega}}|}}\ket{\omega}\otimes\ket{g(\omega)}\;.
\end{equation} By measuring the ancilla qubit and post-selecting the outcome \( \ket{1} \), the remaining system collapses into a quantum state that encodes a uniform superposition over the filtered subset:
\begin{equation}
    \sum_{\omega \in X_{S,\tilde{\omega},g}} \frac{1}{\sqrt{|X_{S,\tilde{\omega},g}|}} \ket{\omega}\;.
\end{equation} 
The probability of successfully measuring the ancilla in state \( \ket{1} \) is given by
\begin{equation}
    p_1 = \frac{|X_{S,\tilde{\omega},g}|}{|X_{S,\tilde{\omega}}|},
\end{equation}
which implies that, on average, the procedure must be repeated \( \frac{|X_{S,\tilde{\omega}}|}{|X_{S,\tilde{\omega},g}|} \) times to obtain a successful outcome. When the size of the selected subset is not exponentially smaller than the full equivalence class, filtering can be carried out efficiently on a quantum computer.

\subsubsection*{Grammar equivalence problem}

A formal grammar can be viewed as a special case of a string rewriting system in which attention is restricted to the equivalence class of strings reachable from a designated starting string $\tilde{\omega}$. The language generated by the grammar is obtained by filtering this equivalence class to retain only those strings that do not contain a designated set of symbols, called non-terminals. This construction reflects the interpretation of grammars as generative mechanisms for syntactically well-formed sentences~\cite{chomsky2002syntactic}. The grammar equivalence problem then consists of determining whether two grammars generate the same language, that is, whether their filtered equivalence class coincide. In our language, if $g$ denotes a Boolean function that selects strings containing no non-terminal symbols, the grammar equivalence problem reduces to estimating the overlap $F$ between the filtered sets $X_{S_1,\tilde{\omega},g}$ and $X_{S_2,\tilde{\omega},g}$, generated by two rewriting systems $S_1$ and $S_2$ acting on the same initial string $\tilde{\omega}$.

\subsubsection*{Estimating classical expectation values}

Finally, a wide range of statistical information can be extracted by measuring the system in the state \( \ket{X_{S,\tilde{\omega}}} \), which results in uniformly sampling strings from the equivalence class \( X_{S,\tilde{\omega}} \). This sampling process enables the estimation of expectation values of classical functions over strings. For example, we can estimate the probability of finding a particular character or substring at a specified position by counting its occurrences in the sampled strings.

\subsubsection*{Estimating quantum expectation values}

We can also estimate expectation values that are not associated with any classical function. Indeed, the expectation value of any Hermitian operator $\langle\hat O\rangle$ that can be expressed as a sum of a polynomial number of tensor product operators can be estimated on a quantum computer via single qubit rotations and sampling. For example, in the case of a binary alphabet (\( |A| = 2 \)), the number of connected strings \( |X_{S,\tilde{\omega}}| \) can be estimated by measuring the expectation value of the observable:
\begin{equation}\label{eq:counting_operator}
    \hat{O} = \left(\hat{\id} + \hat{\sigma}_X\right)^{\otimes L}
\end{equation}
where \( \hat{\sigma}_X \) is the Pauli-X operator and \( L \) is the string length. The operator $\hat{O}$ is the projector onto the equal superposition of all computational basis states, scaled by a factor of $2^L$. Its expectation value yields the fidelity between the orbit state and the uniform superposition state (in Eq.~(\ref{eq:counting_fraction})~), multiplied by the size of the orbit state. This directly provides the total size of the equivalence class, i.e., a solution for the counting problem.

\subsubsection*{Estimating non-linear quantities}

Finally, by preparing multiple copies of the orbit state, one can also estimate non-linear quantities, such as the $2$-Rényi entropy of subsystems~\cite{Islam2015}. This provides insights into the correlations between substrings located in different regions of the string. In particular, entropy serves as an indirect measure of computational complexity: when the Rényi entropy of a region tends toward zero, the structure of the equivalence class simplifies, becoming close to a Cartesian product of independent components. This factorization allows for a compact classical representation of the equivalence class and may lead to significantly faster classical algorithms for many tasks.

\subsection*{Tensor network implementations}

In certain instances, the quantum algorithm introduced in this work admits efficient classical simulation via tensor network (TN) methods~\cite{PhysRevB.94.165116, RevModPhys.77.259, SCHOLLWOCK201196, Montangero2018, Silvi2019}. TN are numerical techniques that express quantum states as networks of tensors with contractions between indices. The size of each tensor index is determined by the local Hilbert space dimension $d$ or by an integer parameter $\chi$, referred to as the \textit{bond dimension}. While a faithful representation of a generic quantum state typically demands a bond dimension that scales exponentially with system size $L$, this requirement can be significantly relaxed when entanglement between subsystems is limited. In such cases, the bond dimension may scale polynomially, enabling efficient classical representations of quantum states and allowing for the simulation of specific processes and measurements. This representation is especially efficient for states obeying an area-law for entanglement, such as ground states of local gapped Hamiltonians~\cite{RevModPhys.82.277, Evenbly2011}.

A Matrix Product State (MPS) is a one-dimensional instance of a tensor network, which represents a one-dimensional lattice of $L$ qudits using $L$ rank-3 tensors of shape $(\chi, d, \chi)$. The memory cost to store an MPS scales as $\mathcal{O}(L \cdot \chi^2 \cdot d)$. We simulate the time evolution within the MPS formalism using a Time-Dependent Variational Principle (TDVP)~\cite{PhysRevB.94.165116}. A variety of relevant quantities — including fidelities, expectation values of observables, sampling, and subsystem entropies — can be efficiently computed, provided that $\chi$ grows sub-exponentially with system size~\cite{Montangero2018}.

When the entanglement of the orbit state admits an efficient tensor network representation, the performance of the quantum algorithm can be assessed through classical simulation. In particular, the preparation of orbit states via imaginary quantum annealing is especially well-suited to tensor network methods, as it naturally suppresses excitations introduced by numerical errors and by approximations associated with a finite bond dimension.

Beyond serving as a testbed for the quantum normal form reduction algorithm, the tensor network representation naturally gives rise to a quantum-inspired algorithm. In the current computational landscape — where quantum hardware remains in its early stages while classical computing is highly mature — such an approach offers a novel and potentially advantageous framework for addressing the challenges inherent in term rewriting. Additionally, tensor networks enable the efficient computation of properties of quantum states that are otherwise difficult, or even exponentially hard, to extract on quantum hardware. Notable examples include the evaluation of von Neumann entropies and the estimation of exponentially small probability amplitudes~\cite{Ballarin2024}.

The primary limitation of the tensor network approach proposed here lies in the complexity of local correlations among strings within the equivalence class $X_{S,\omega}$. These correlations can become particularly intricate in certain relevant contexts, for example, in natural languages, where meaning emerges from the nuanced interplay among components of a sequence~\cite{e19070299, PhysRevA.111.032409, Gallego_2022}. In this case, more sophisticated TN structures, such as Tree Tensor Networks~\cite{PhysRevA.74.022320,PhysRevA.81.062335,PhysRevB.90.125154}, could allow for an accurate simulation. An even more challenging scenario arises in rewriting systems that go beyond string rewriting. In one-dimensional string rewriting systems, the locality of the rules translates into a one-dimensional local structure for the Laplacian, which can yield a relatively simple entanglement pattern in the ground states and the possibility to approximate these states as MPS with bond dimension that scales polynomially with the system size.~\cite{arad2013arealawsubexponentialalgorithm}. One-dimensional string rewriting appears in practice in DNA mutation modeling, formal languages, regex-based text processing, compiler peephole optimization, and grammar-based data compression. In contrast, when rewriting rules are non-local, such as large tree-level transformations used in compilers and symbolic algebra, the notion of one-dimensional distance between subsystems breaks down, making an efficient tensor network representation more challenging or, in some cases, inefficient. Nevertheless, recent advances in tensor network algorithms have demonstrated the ability to represent ground states of systems beyond one dimension through Projected Entangled Pair States~\cite{Evenbly2011}, Multi-scale Entanglement Renormalization Ansatz~\cite{Evenbly2011}, and Augmented Tree Tensor Networks~\cite{Felser_2021}.

\subsection*{Numerical results}

We simulate the quantum algorithm using tensor network methods and observe a favorable scaling of the required computational resources with the size of the input strings. 


We consider the string rewriting system $S= \Big[ A | R \Big]$, where $A=\{a,b\}$ and 
\begin{align}\label{eq:srs_def}
    R = & \{\nonumber a_{[i]}a_{[i+1]}a_{[i+2]}a_{[i+3]} \approx b_{[i]}a_{[i+1]}b_{[i+2]}a_{[i+3]},\nonumber\\
    &b_{[i]}a_{[i+1]}b_{[i+2]}a_{[i+3]} \approx b_{[i]}b_{[i+1]}b_{[i+2]}b_{[i+3]}\}\;.
\end{align}
Despite its simplicity, the application of the aforementioned rewriting system to a string of length $L$ generates an exponential number of connected strings. We validate the quantum framework introduced in this work by solving a collection of instances of the word problem and the counting problem for the SRS defined in Eq.~(\ref{eq:srs_def}). Each problem instance is specified by two disjoint sets of four strings of length $L$:  
\begin{align}
    \Omega_1(L) &= \{\omega_1, \omega_2, \omega_3, \omega_4\}\nonumber\\
    \Omega_2(L) &= \{\omega_5, \omega_6, \omega_7, \omega_8\}\;.
\end{align}
Within each set, all pairs of strings are mutually connected through a sequence of rewriting operations, i.e.,
\begin{equation}
    \Omega_1(L) \subset X_{S,\omega_1} = X_{S,\omega_2} = X_{S,\omega_3} = X_{S,\omega_4}\;,
\end{equation}
and
\begin{equation}
\Omega_2(L) \subset X_{S,\omega_5} = X_{S,\omega_6} = X_{S,\omega_7} = X_{S,\omega_8}\;.    
\end{equation}
No string from one set is connected to any string in the other, which guarantees that the corresponding equivalence classes are distinct, for example \( X_{S,\omega_1} \neq X_{S,\omega_5} \).

We consider problem instances with string lengths \( L \) ranging from \( 10 \) to \( 100 \). For each string $\tilde{\omega}$ belonging to $\Omega_1(L)$ or to $\Omega_2(L)$, we construct the corresponding orbit state by simulating the time evolution governed by the imaginary-time Hamiltonian:
\begin{equation}
    \hat{I}_{S,\tilde{\omega}}(h) = i\left[\frac{t}{\tau}\hat{\mathcal{L}}_S - \left(1-\frac{t}{\tau}\right) \ket{\tilde{\omega}}\bra{\tilde{\omega}}\right]
\end{equation}
for a long enough annealing time~$\tau$. Here, \( \hat{\mathcal{L}}_S \) denotes the Laplacian operator associated with the rewriting system \( S \). 

To assess the quality of the prepared ground states, we evaluate both their flatness, quantified by the Dirichlet energy $E_\text{D} = \langle \psi|\hat{\mathcal{L}}_S|\psi\rangle$, and their non-connected probability $p_\mathrm{NC}$, defined as the probability of sampling a string which is not connected to the input string when performing measurements in the computational basis, that is, an error in our classification.

\begin{figure}[t!]
\includegraphics[width=\linewidth]{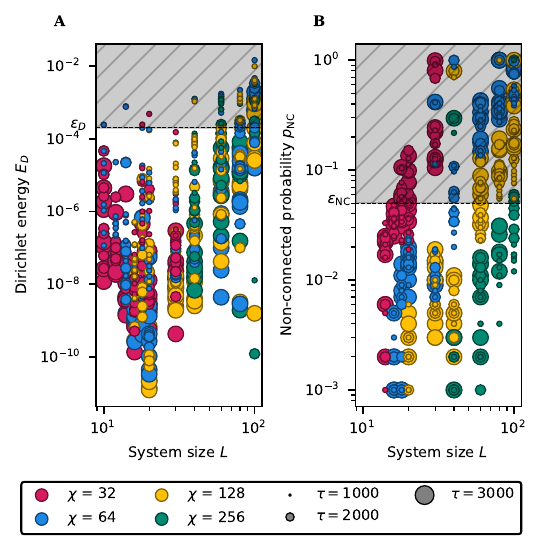}
\caption{
\textbf{Quality of the orbit state MPS obtained via imaginary-time evolution as a function of the system size $L$, for different input string $\omega$, bond dimensions $\chi$, and annealing times $\tau$.} \textit{Panel~A:} Dirichlet energy $E_D$ of the state.  We accept only states with energy below the tolerance $\epsilon_\text{D}$. The energy tends to decrease with increasing annealing time. \textit{Panel~B:} Probability $p_\mathrm{NC}$ of sampling a string which is not connected to $\omega$. We accept only states with probability below the tolerance $\epsilon_\mathrm{NC}$. The probability tends to decrease with increasing bond dimension. The probability $p_\mathrm{NC}$ is estimated by performing $n_S = 1000$ samples in the computational basis~\cite{Ballarin2024}. Separately, we apply the Knuth-Bendix algorithm to verify how many sampled strings are connected to the input string.}
\label{fig:tn_quality}
\end{figure}

In Figure~\ref{fig:tn_quality}, we present the behavior of the Dirichlet energy $E_D$ and the non-connected probability $p_\mathrm{NC}$ as a function of the system size. We consider different input strings $\omega$, as well as varying bond dimensions $\chi$ and annealing times $\tau$. As expected, increasing the annealing time systematically lowers the Dirichlet energy of the final state. When the bond dimension is sufficiently large, the non-connected probability also decreases toward zero, indicating that the evolution converges to the orbit state. Conversely, if the bond dimension is too small, the projection onto the MPS manifold can steer the evolution toward alternative ground states of the Laplacian, which are typically linear combinations of orbit states and exhibit a high non-connected probability $p_\mathrm{NC}$. An extreme example of this behavior is the equal-amplitude superposition of all computational basis states, which is a ground state of the Laplacian and can be exactly represented with bond dimension $\chi = 1$.

\begin{figure}[t!]
\includegraphics[width=\linewidth]{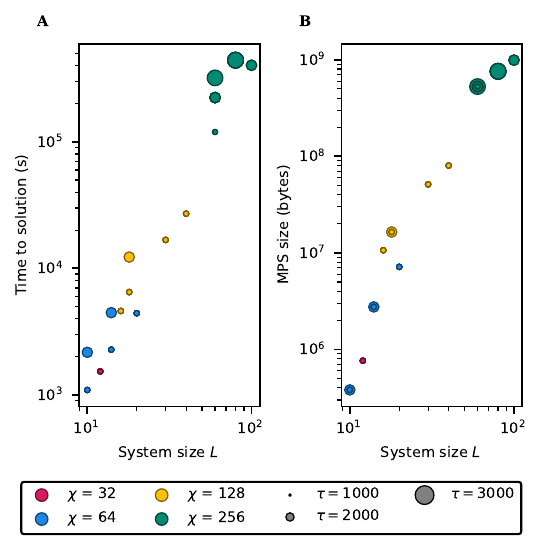}
\caption{\textbf{Computational resources required to construct and store orbit states as MPS.} Distinct points with the same value of $L$ correspond to different input strings. \textit{Panel~A:} Computational time required to simulate the imaginary-time evolution.  This time is governed by two main factors: first, the cost of simulating a single time step of the evolution via TDVP, which scales asymptotically as $\mathcal{O}(\chi^3 L d^2)$; and second, the number of time steps needed to reach a sufficiently accurate approximation of the orbit state. The latter is determined by the annealing time $\tau$, which controls the convergence rate of the IQA process and also provides an estimate of the computational cost when the algorithm is implemented on quantum hardware. In turn, $\tau$ is governed by the spectral properties of the Hamiltonian $\hat{H}_{S,\tilde{\omega}}$, including its gap and low-energy structure, and is therefore ultimately dictated by the specific rewriting system under consideration.
\textit{Panel~B:} Memory footprint of the MPS representation of the orbit state as a binary file, which scales asymptotically as $\mathcal{O}(\chi^2 L d)$ with the input size.}
\label{fig:tn_resources}
\end{figure}

\begin{figure}[t!]
\includegraphics[width=\linewidth]{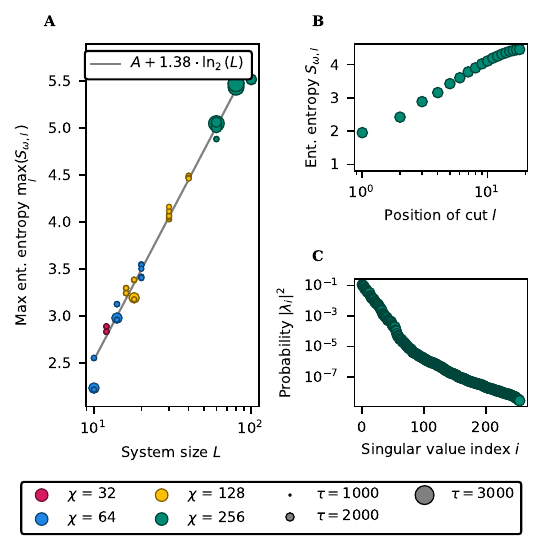}
\caption{\textbf{Entanglement of orbit states.} \textit{Panel~A:} Base $2$ entanglement entropy maximized over all the bipartitions, as a function of the system size $L$, for the MPS generated in the shortest time among those that exceed the quality tolerance. In an open one-dimensional chain, the entanglement between two adjacent regions can be quantified by the entropy of the squared Schmidt coefficients $|\lambda_i|^2$ obtained from a bipartition at position $l$. \textit{Panel~B:} Entanglement entropy as a function of the position of the cut for MPS representing an orbit state of $40$ qubits. \textit{Panel~C:} Entanglement spectrum of the largest bipartition, for an orbit state of $40$ qubits.}
\label{fig:tn_entanglement}
\end{figure}

\begin{figure}[t!]
\includegraphics[width=\linewidth]{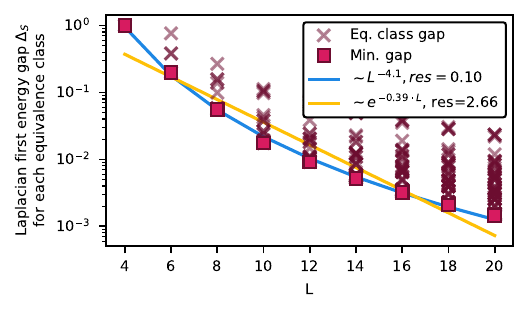}
\caption{\textbf{Polynomial scaling of the first energy gaps of the Laplacian $\hat{\mathcal{L}}_S$ for the string rewriting system in the exam.} The cross markers indicate gaps for the Laplacian restricted to each possible equivalence class, while the square marker indicates the minimum gap among all the equivalence classes. The minimum gap is compared with the corresponding polynomial and exponential least square fits. \textit{res} is the sum of squared residuals. The polynomial fit yields smaller residuals, lending stronger support to a polynomial trend in the scaling of the gap—and consequently, in the simulation time.
}
\label{fig:first_gap}
\end{figure}

\begin{figure}[t!]
\includegraphics[width=\linewidth]{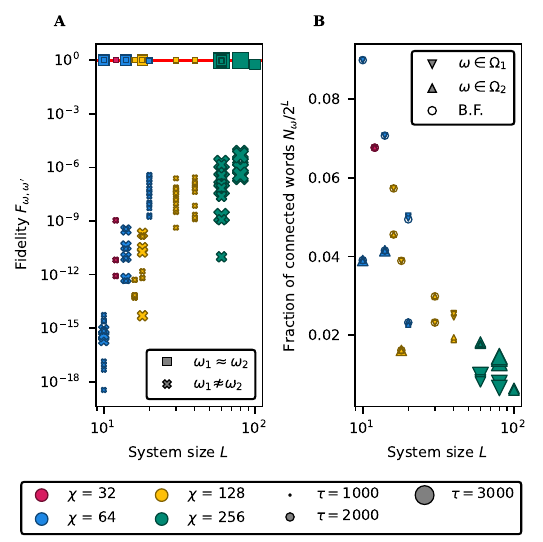}
\caption{\textbf{Solution of the word problem and counting problem for different system sizes $L$, bond dimensions $\chi$, and annealing times $\tau$.}
\textit{Panel~A:} Fidelity between pairs of orbit states. The marker shape distinguishes between pairs of connected and disconnected words. Orbit states associated with connected words exhibit fidelities close to $1$, while those corresponding to disconnected words exhibit fidelities close to $0$.
\textit{Panel~B:} Estimated number $N_\omega$ of connected words for each input string, computed as the expectation value of the operator $\hat O$ in Eq.~(\ref{eq:counting_operator}) on the corresponding orbit state, and normalized with respect to the size of Hilbert space $2^L$. The triangular marker shapes distinguish between words belonging to different disconnected sets. The circular markers represent results obtained via exhaustive graph exploration. The presence of non-overlapping triangular markers for different input words reflects the uncertainty arising from the finite quality tolerance used to select converged MPS.}
\label{fig:tn_word_and_count}
\end{figure}

For the remainder of this analysis, we consider only MPS with both low non-connected probability ($p_\mathrm{NC} \leq \varepsilon_{NC} = 0.05$) and low Dirichlet energy ($E_D \leq \varepsilon_D = 0.0002$). We select the MPS generated in the shortest time among those that exceed the quality tolerance. This selection allows us to estimate the minimal computational time and memory required to classically encode orbit states via TN. As the cost of computing fidelities and expectation values is negligible compared to that of state preparation, the total runtime can be interpreted as the time-to-solution for both the word and counting problems.

The computational resources required to prepare orbit states via TN techniques are summarized in  Figure~\ref{fig:tn_resources}. In the figure, we observe that increasing the annealing time from $\tau = 1000$ to $\tau = 3000$ suffices to scale the computation from systems of length $L = 10$ to $L = 100$. This trend suggests that the annealing time $\tau$ grows slowly with system size, supporting the possibility of efficient scaling and motivating further exploration toward a physical implementation of the algorithm. In the same regime, the bond dimension required to approximate orbit states above the quality threshold increases from $\chi = 32$ to $\chi = 256$.

Although the computational costs shown in Figure~\ref{fig:tn_resources} are consistent with polynomial scaling, our analysis is currently limited to systems of size up to $L = 100$, constrained by the computational effort required to simulate larger instances. As a result, Figure~\ref{fig:tn_resources} alone may not fully capture the asymptotic behavior of the algorithm. To strengthen the case for a polynomial scaling for the string rewriting system in the exam, we turn to the entanglement structure of the orbit states, which directly impacts the bond dimension $\chi$ required for an accurate MPS approximation. As shown in Figure~\ref{fig:tn_entanglement}, the maximum entanglement entropy across all bipartitions increases logarithmically with both the system size (Panel~A) and the size of the largest bipartition of the system (Panel~B). Moreover, for each bipartition, the Schmidt singular values exhibit an exponential decay (Panel~C). Together, these observations imply that the number of singular values required to achieve a high-fidelity approximation grows only polynomially with the bipartition size—and hence with the overall system size. This favorable entanglement structure supports the efficient representability of orbit states using MPS with a bond dimension that scales polynomially.

Since the Laplacian inherits locality from the rewriting rules, the observed polynomial scaling of annealing time, bond dimension, and computational cost can be deduced by the scaling of the spectral gap of the Laplacian. Indeed, as shown in the Section \textit{Computational complexity and final energy gap} of the Supplementary Text, the number of time steps required to approximate the orbit state associated with an input word $\tilde{\omega}$ scales as $\mathcal{O}(\Delta_{\tilde{\omega}}^{-2})$, where $\Delta_{\tilde{\omega}}$ is the smallest nonzero eigenvalue of the Laplacian restricted to the corresponding equivalence class $X_{S,\tilde\omega}$. Each time step is simulated via a time-dependent variational principle algorithm that has computational complexity $\mathcal{O}(\chi^3)$ and memory complexity $\mathcal{O}(\chi^2)$. We also show that the errors accumulated during the imaginary-time annealing process are exponentially suppressed, implying that the bond dimension $\chi$ needed to reach good accuracy is mainly determined by the final stage of the evolution, i.e., in the vicinity of the orbit state. Since the Laplacian is a local operator, the bond dimension required to approximate the orbit states scales polynomially in the inverse gap, more precisely as $\mathcal{O}(\Delta_{\tilde{\omega}}^{-1/3})$~\cite{arad2013arealawsubexponentialalgorithm}. In this way, the overall computational complexity of the proposed algorithm is controlled by the final spectral gap $\Delta_{\tilde{\omega}}$. The value of this gap depends on the structure of the underlying rewriting system. For the rewriting system considered in our example, Figure~\ref{fig:first_gap} reports the gaps of the Laplacian restricted to each equivalence class, for system sizes up to $20$. These gaps govern the average-case complexity when the input word is sampled uniformly at random. The smallest gap scales polynomially with system size, implying polynomial worst-case computational complexity.

Once orbit states are available, measuring fidelities and observables enables efficient solutions to both the word problem and the counting problem. The results of these tasks are presented in Figure~\ref{fig:tn_word_and_count}.

\subsubsection*{Word problem}

In Figure~\ref{fig:tn_word_and_count} Panel~A), we illustrate the solution of the word problem via fidelity measurements between orbit states. Pairs of words that belong to the same equivalence class correspond to orbit states with high fidelity, whereas disconnected word pairs exhibit low fidelity. These results demonstrate that our algorithm correctly solves the word problem. Greedy strategies based on explicit graph exploration, such as breadth-first search, require exponential time and cannot reach comparable system sizes. In contrast, the Knuth-Bendix completion algorithm solves the word problem in polynomial time and remains faster within the system sizes explored here (see Supplementary Text). However, a definitive performance comparison would require extending the analysis to larger system sizes and a broader range of rewriting systems. 

\subsubsection*{Counting}

In Figure~\ref{fig:tn_word_and_count} Panel~B), we estimate the number of connected words for each input word. Words belonging to the same set are associated with the same number of connected words. We reconstruct connected sets containing up to $10^{28}$ words of length $100$. Storing this amount of information sequentially would require approximately $10^{17}$ terabytes of memory, underscoring the compression power of the quantum-inspired representation. For comparison, we also report the exact counts obtained via greedy enumeration for $L \leq 30$, which agree with our estimates. This baseline was computed using greedy graph exploration, i.e., breadth-first search with memoization~\cite{10.5555/1614191}, which constitutes the state-of-the-art alternative for exact enumeration. Its computational cost grows polynomially with the connected set size—and hence exponentially with $L$—making it unfeasible for large-scale instances. In contrast, our quantum normal form reduction technique enables the enumeration of connected sets at scales inaccessible to graph exploration.

\section*{Discussion}

We have introduced quantum normal form reduction, a general paradigm for automating equational reasoning on quantum computers. Our approach leverages the ability of quantum systems to encode and manipulate exponentially large sets of semantically equivalent symbolic expressions as a single quantum state, an \emph{orbit state}. Orbit states are prepared as the ground states of suitable sparse Hamiltonians, which can be efficiently simulated on quantum devices when the rewriting system encoding the equivalence relations contains a polynomial number of rules in the string size $L$. Quantum optimization techniques are employed to prepare orbit states. As it is common in many optimization problems, the computational cost of these procedures depends on the specific problem instance.

We have simulated our algorithm using tensor network techniques, demonstrating its effectiveness in solving the word problem and the counting problem for a toy rewriting system. The results obtained from these simulations suggest the potential for a quantum advantage in equational reasoning, as well as for the development of novel quantum-inspired algorithms capable of outperforming classical approaches. While our results are promising, the present TN emulations are currently limited to strings up to $100$ characters.  This limitation primarily stems from the growth in classical computational time associated with the increasing entanglement of orbit states, which appears to scale polynomially within the investigated size range. Further investigation is necessary to corroborate any general claim regarding quantum or quantum-inspired speedups.  Interestingly, tensor network algorithms have been shown to outperform state-of-the-art solvers for a related counting problem, namely counting the solutions of SAT problems~\cite{Biamonte_2015, 10.21468/SciPostPhys.7.5.060}. Potential connections between these algorithms and the methods introduced in this work remain to be explored. The comparison with state-of-the-art classical algorithms (see Supplementary text) clarifies in which limits quantum normal form reduction can be regarded as a quantum extension of classical
methods, and how insights from these approaches may be leveraged to improve the proposed quantum approach.

The tensor-network approach demonstrates the possibility of using computational tools from many-body physics in equational reasoning. For instance, one may employ density matrix renormalization group methods~\cite{RevModPhys.77.259} to approximate a random ground state of the Laplacian operator, i.e., a random superposition of orbit states. The fidelity between an input word and a random ground state obtained in this way would define a sound and complete equational hash, i.e., a function constant within each equivalence class but different across distinct classes. Once constructed, this function allows one to solve the word problem for any pair of words simply by comparing their hash values, without explicitly reconstructing the corresponding orbit states.

Among the many potential real-world applications outlined in the introduction, the design of quantum algorithms for formal language processing represents a particularly significant future direction. Since formal grammars capture the underlying structure of both human and programming languages, this could open new perspectives for the development of quantum algorithms in language processing~\cite{wiebe2019quantumlanguageprocessing, coecke2020foundationsneartermquantumnatural} and software design. Moreover, grammar-based algorithms~\cite{grammar_based_codes} for lossless compression of classical data could be further improved through the quantum techniques introduced here.

Another promising future direction is the development of quantum algorithms for optimizing cost functions within an equivalence class. A preliminary example is the design of quantum algorithms for quantum circuit compilation~\cite{rattacaso2024quantumcircuitcompilationquantum}, where the set of circuits implementing the same unitary operator is explored through quantum dynamics generated by a set of Hermitian operators that encode rewriting rules.

Finally, our findings underscore the effectiveness of TN as a compressed representation for extensive datasets structured by equivalence rules, a property of increasing relevance in the era of massive data generation, and suggest promising directions toward real-world applications.

Beyond tensor network simulation, the proposed algorithm can also be implemented using quantum circuits (see Supplementary text). This circuit-based implementation relies on multi-controlled gates, which can be efficiently realized on universal quantum computers using a linear number of elementary gates and ancilla qubits~\cite{PhysRevA.52.3457}. Moreover, such gates may be natively supported on certain hardware platforms, including Rydberg-atom arrays and superconducting circuits~\cite{PhysRevX.10.021054}.

\section*{Materials and Methods}

\subsection*{The graph Laplacian operator}

Here we construct the graph Laplacian operator $\hat{\mathcal{L}}_S$ whose expectation value corresponds to the Dirichlet energy in Eq.~(\ref{eq:dirichelet}).

First, we define the discrete Laplacian matrix
\begin{equation}\label{eq:lap_matix}
\mathcal{L}_{kk'}=
    \begin{cases}
		\deg(\omega_k), & \text{if } k=k'\\
        -\text{mult}(\omega_k,\omega_{k'}), & \text{if } k\neq k'\,;
	\end{cases}
\end{equation}
where the degree $\deg(\omega_k)$ is the number of edges attached to the vertex $\omega_k$, and the multiplicity $\text{mult}(\omega_k,\omega_{k'})$ is the number of edges connecting $\omega_k$ and $\omega_{k'}$. The Dirichlet energy of $\psi$ can be expressed via the Laplacian matrix as
\begin{equation}
    E_\text{D} = \sum_{k,k'}\mathcal{L}_{kk'}\psi_{k}^*\psi_{k'}\;.
\end{equation}
By introducing the Laplacian operator
\begin{equation}
    \hat{\mathcal{L}}_S = \sum_{kk'}\mathcal{L}_{kk'}\ket{\omega_k}\bra{\omega_{k'}}\;,
\end{equation}
we can finally express the Dirichlet energy as the expectation value of $\hat{\mathcal{L}}_S$ on the quantum state $\ket{\psi}$:
\begin{equation}
    E_\text{D} = \langle \psi|\hat{\mathcal{L}}_S|\psi\rangle\;.
\end{equation}
By construction, the Dirichlet energy is always larger than or equal to zero, and is zero only for a function that is constant on each subgraph (see Eq.~(\ref{eq:dirichelet}) ). Thus, the Laplacian operator is a positive semi-definite operator having the orbit states as degenerate ground states with zero energy.

As the graph $G$ is generated by the action of the rewriting rules, the Laplacian operator is a function of the rewriting rules in $R$. In particular, the entire Laplacian operator is the sum of Laplacian operators $\hat{\mathcal{L}}_r$ of the graphs induced by each single rule $\hat{r}$, since the degeneracy of a vertex is the sum of the degeneracies introduced by each rule, and the multiplicity of an edge is the sum of multiplicities. The Laplacian induced by the rule $r$ can be written as
\begin{align}
\hat{\mathcal{L}}_r &= \hat{r}^2 - \hat{r} \nonumber\\
&=\Big(\ket{\alpha}\bra{\alpha}_{j_1}\otimes\ket{\beta}\bra{\beta}_{j_2}\otimes\dots+\ket{\alpha'}\bra{\alpha'}_{j_1}\otimes\ket{\beta'}\bra{\beta'}_{j_2}\nonumber\\&-\ket{\alpha'}\bra{\alpha}_{j_1}\otimes\ket{\beta'}\bra{\beta}_{j_2}\otimes\dots-\ket{\alpha'}\bra{\alpha}_{j_1}\otimes\ket{\beta'}\bra{\beta}_{j_2}\Big)\;.
\end{align}
In this expression, the first two terms are diagonal in the computational basis and, for each basis state, count the number of configurations connected via rule $r$, thereby contributing to the vertex degrees in Eq.~(\ref{eq:lap_matix}). The last two terms contribute to the off-diagonal structure of Eq.~(\ref{eq:lap_matix}), assigning a weight of $1$ to pairs of basis states connected by rule $r$ and $0$ otherwise. Finally, we write the Laplacian operator of the whole rewriting system as the sum of the Laplacians of each rule:
\begin{equation}
    \hat{\mathcal{L}}_S= \sum_{r\in R} \hat{\mathcal{L}}_r = \sum_{r\in R}\left( \hat{r}^2 - \hat{r}\right)\;.
\end{equation}

The ground state of $\hat{\mathcal{L}}_S$ is also a ground state for each rule-associated operator $\hat{\mathcal{L}}_r$, so that $\hat{\mathcal{L}}_S$ is frustration-free.

\subsection*{Instances generation for the numerical experiment}

In the numerical experiment, we consider different instances for the word problem and counting. Each instance consists of two disjoint sets of four strings of length $L$. Pairs of strings within the same set are connected through some sequence of rewriting operations, while no string from one set is connected to any string in the other. Each set of connected strings is generated by applying the Knuth-Bendix algorithm to reduce a randomly sampled initial string to distinct but connected strings, i.e., the normal forms produced by the Knuth-Bendix algorithms for different orderings (see Supplementary Text for details about normal forms and the Knuth-Bendix algorithms). We use the computer algebra system \textit{GAP}~\cite{GAP4} to run the Knuth-Bendix algorithm. Additional random applications of the rewriting rules in $S$ are then performed to diversify the strings within each equivalence class. The construction of normal forms also allowed us to verify that no string in $\Omega_1$ is connected to any string in $\Omega_2$.

\subsection*{Simulation details}

The imaginary quantum annealing evolution is simulated using the MPS formalism and a Time-Dependent Variational Principle (TDVP)~\cite{PhysRevB.94.165116} via the tensor network emulator \textit{Quantum TEA Leaves}~\cite{qtealeaves}.

The Hamiltonian dynamics are discretized in time steps of size \( \delta_t = 0.5 \). Note that the Trotter error introduced by $\delta_t$ decreases with the annealing time $\tau$. Indeed, increasing $\tau$ reduces the change in the Hamiltonian at each time step, thereby decreasing the magnitude of the commutator term arising from the Baker–Campbell–Hausdorff formula. For each input string, we perform simulations for annealing times \( \tau \in \{1000, 2000, 3000\} \) and bond dimensions \( \chi \in \{32, 64, 128, 256\} \). Increasing $\tau$ and $\chi$ enhances the fidelity of the resulting orbit state with respect to the ideal target, but at the cost of increased computational resources.

All simulations were performed on a virtual machine equipped with 20 AMD EPYC 7413 CPUs and 128 GB of memory, using parallel execution across groups of eight input strings with identical length, bond dimension, and annealing time.


\section*{Acknowledgments}
We acknowledge Nicola Assolini, Diego di Bernardo, Alessandra Di Pierro, Aleks Kissinger, Sergii Strelchuk and David Yu Yuan for useful discussions and valuable feedback.
\paragraph*{Author Contributions:}
Conceptualization: DR, MB, IS, DJ, SM. Methodology: DR, MB, DJ. Software: DR, DJ. Validation: DR. Formal analysis: DR. Investigation: DR. Resources: DJ. Data curation: DR. Writing - original draft: DR. Writing - review \& editing: MB, IS, DJ, SM. Visualization: DR. Supervision: IS, DJ, SM. Project administration: IS, SM. Funding acquisition: SM.

\paragraph*{Funding:}
The research leading to these results has received funding from the following organizations: European Union via Italian Research Center on HPC, Big Data and Quantum Computing (NextGenerationEU Project No. CN00000013), project EuRyQa (Horizon 2020), project PASQuanS2 (Quantum Technologies Flagship); Italian Ministry of University and Research (MUR) via: Quantum Frontiers (the Departments of Excellence 2023-2027); the World Class Research Infrastructure - Quantum Computing and Simulation Center (QCSC) of Padova University; Istituto Nazionale di Fisica Nucleare (INFN): iniziativa specifica IS-QUANTUM; the German Federal Ministry of Education and Research (BMBF) via the project QRydDemo. We acknowledge computational resources from Cloud Veneto, as well as computation time on Cineca’s Leonardo machine.
\paragraph*{Competing interests:}
The authors declare that they have no competing interests.
\paragraph*{Data and materials availability:}
Data, software, simulation outputs, and all instructions needed to reproduce the results of this paper are available on Zenodo~\cite{datasets_and_figures}.


\subsection*{Supplementary materials}
Supplementary Text\\

\bibliography{refs}
\bibliographystyle{sciencemag}

\clearpage
\onecolumn

\renewcommand{\thefigure}{S\arabic{figure}}
\renewcommand{\thetable}{S\arabic{table}}
\renewcommand{\theequation}{S\arabic{equation}}
\renewcommand{\thepage}{S\arabic{page}}
\setcounter{figure}{0}
\setcounter{table}{0}
\setcounter{equation}{0}
\setcounter{page}{1} 

\section*{Supplementary Materials for\\ \scititle}

Davide~Rattacaso$^{1,2\ast}$,
Daniel~Jaschke$^{1,2,3,4}$,
Marco~Ballarin$^{1,2,5}$,\\
Ilaria~Siloi$^{1,2}$,
Simone~Montangero$^{1,2}$\\
\small$^{1}$Dipartimento di Fisica e Astronomia “G. Galilei” \& Padua Quantum Technologies Research Center,\\ \small Università degli Studi di Padova, Italy I-35131, Padova, Italy.\\
\small$^{2}$INFN, Sezione di Padova, via Marzolo 8, I-35131, Padova.\\
\small$^{3}$Institute for Complex Quantum Systems, Ulm University, Albert-Einstein-Allee 11, 89069 Ulm, Germany.\\
\small$^{4}$Current affiliation: PlanQC GmbH, Lichtenbergstr. 8, 85748 Garching, Germany.\\
\small$^{5}$Current affiliation: Quantinuum, Partnership House, Carlisle Place, London SW1P 1BX, United Kingdom.\\
\small$^\ast$Corresponding author. Email: davide.rattacaso@unipd.it

\subsection*{Simulating the Hamiltonian}

Here, we analyze the quantum resources required for simulating the Hamiltonian $\hat H_{S,\tilde{\omega}}$, specifically in terms of the number of quantum gates.

We do not consider native qudit platforms~\cite{10.3389/fphy.2020.589504}, which might eventually prove advantageous for operating with rewriting systems whose alphabet contains more than two elements. Instead, we focus on standard qubit-based computation. To this end, we note that any rewriting system defined on an alphabet of size $d$ can be translated into a rewriting system over a binary alphabet, as happens when strings are manipulated in standard (Boolean) classical computers. In particular, each symbol of the original alphabet can be encoded as a binary string of length $\left\lceil \log_2{d} \right\rceil $, i.e., its binary logarithm rounded to the nearest bigger integer. This corresponds to the binary representation of the symbol index. For example, for the alphabet $A = \{a, b, c\}$, we can assign: $a \rightarrow (0,0)$, $b \rightarrow (0,1)$, and $c \rightarrow (1,0)$. Similarly, a string of $L$ characters over $A$ is mapped to a binary string of length $L \cdot \left\lceil \log_2{d} \right\rceil$. Each rewriting rule of length $l$ is likewise mapped to a rule of length $l \cdot \left\lceil \log_2{d} \right\rceil $, while the total number $n_r$ of rewriting rules remains unchanged.

With this binary encoding in place, we can restrict our analysis to simulating $\hat H_{S,\tilde{\omega}}$ over the binary alphabet $A = \{0,1\}$, which is compatible with standard qubit-based quantum devices.

The simulation typically relies on a discretization of the time evolution, allowing the approximation of the evolution operator as a product of exponentials of the individual Hamiltonian terms~\cite{doi:10.1126/science.273.5278.1073}. The Trotterized evolution, or any analogous optimization ansatz such as QAOA, can be implemented by the application of the operator
\begin{equation}\label{eq:trotter_step}
U = \prod_i^{N} \left( \mathrm{e}^{i\alpha_i \ket{\tilde{\omega}}\bra{\tilde{\omega}}} \prod_{r \in R} \mathrm{e}^{-i\beta_{r,i} \hat{r}^2} \mathrm{e}^{i\beta_{r,i} \hat{r}} \right)\;,
\end{equation}
where the $\alpha_i$ and $\beta_i$ are real coefficients scaling as the time step $\delta_\tau = \tau/N$, the indices $i$ run over the $N$ time steps, and $\tau$ is the total evolution time. The operator consists of one term involving the exponential of the projector $\ket{\tilde{\omega}}\bra{\tilde{\omega}}$, and a sequence of terms derived from the Laplacian. The number of Laplacian-related operators is twice the number $n_r$ of rewriting rules in the system, due to the squared and linear terms for each rule.

Thus, the entire evolution can be realized by implementing as a circuit consisting of $N\cdot(2n_r+1)$ operators
\begin{equation}\label{eq:expop}
    W = \mathrm{e}^{-i\theta \ket{\mathbf{b''}}\bra{\mathbf{b'}}}\;,
\end{equation}
where $\mathbf{b'}$ and $\mathbf{b''}$ are binary substrings of size $l$ (for operators encoding rules) or $L$ (for the operator encoding the projection on the word $\tilde{\omega}$), and $N$ is the total number of time steps.

\begin{figure}[t!]
\includegraphics[width=0.5\linewidth]{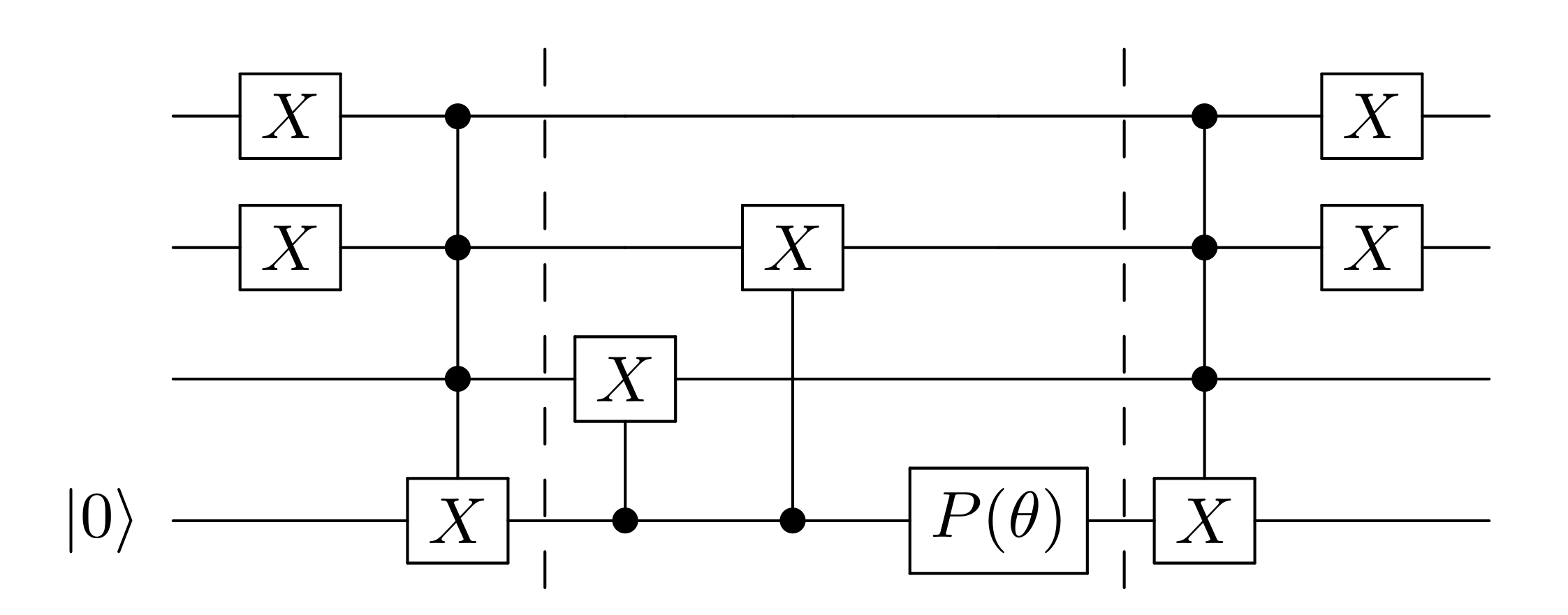}
\caption{\textbf{Circuit implementation of $\mathrm{e}^{-i\theta \ket{010}\bra{001}}$.} In the first part of the circuit, we flip the ancilla only if the state is $\ket{001}$. In the second part, we map $\ket{001}$ to $\ket{010}$ and add a global phase if the ancilla is in the state $\ket{1}$. Finally, we uncompute the ancilla.}
\label{fig:w_circuit}
\end{figure}

The operators in Eq.~(\ref{eq:expop}) can be implemented in a quantum circuit (see Figure~\ref{fig:w_circuit}). First, we use a combination of X gates and a multi-controlled NOT gate to flip an ancilla qubit, conditioned on the local quantum state matching $\ket{\mathbf{b'}}$. Next, we apply two-qubit gates with the control qubit being the ancilla to transform the state $\ket{\mathbf{b'}}$ into $\mathrm{e}^{-i\theta} \ket{\mathbf{b''}}$. Finally, we reverse the ancilla operation by applying the same $X$ gates and multi-controlled NOT gate, effectively uncomputing the ancilla. Thus, when the operator $ W$ to be implemented corresponds to a rewriting rule $ r$, it requires $\mathcal{O}(w_r)$ gates, where $w_r$ denotes the number of characters the rule acts upon. The implementation involves two multi-controlled gates. Similarly, if $W$ represents the evolution generated by the projection onto the input state, it requires $\mathcal{O}(L)$ gates, with $L$ being the system size. In this case as well, only two multi-controlled gates are needed. Multi-controlled gates can be implemented efficiently on universal quantum computers using a linear number of elementary gates and ancilla qubits~\cite{PhysRevA.52.3457}. Alternatively, they may be natively supported on certain hardware platforms, such as Rydberg-atom arrays and superconducting circuits~\cite{PhysRevX.10.021054}. In both scenarios, implementing the operator $W$ requires a number of long-range two-qubit gates that scales linearly with the length of the binary string $\mathbf{b'}$.

Overall, simulating the Trotterized evolution over $N$ steps requires
\begin{equation}
    D = \mathcal{O}(N (2w\cdot n_r + L))
\end{equation}
gates, where $w$ is the maximum number of characters affected by any rule in the rewriting system. These resources are linear in the number of bits needed to describe the problem instance classically—that is, the input string and the set of rewriting rules.

As a consequence of the Baker–Campbell–Hausdorff formula, the first-order Trotterized simulation in Eq.~(\ref{eq:trotter_step}) incurs, at each time step, a local error scaling as $\delta_\tau^2 = \mathcal{O}(\tau^2/N^2)$. Higher-order Trotter–Suzuki decompositions can be employed to systematically reduce this error. Restricting here to the first-order approximation, the cumulative error over the full time evolution scales as $\mathcal{O}(\tau^2/N)$. Therefore, in order to approximate the continuous-time dynamics within a target Trotter error $\epsilon$, one requires $N = \mathcal{O}(\tau^2/\epsilon)$ discrete time steps.

When the objective is ground-state preparation, and assuming a simple adiabatic evolution as the heuristic method of choice, the total evolution time required to reach the target ground state is expected to scale as $\Delta_{\min}^{-2}$~\cite{aqc_review}, where $\Delta_{\min}$ denotes the minimum energy gap of the Hamiltonian restricted to the dynamically accessible subspace. This subspace is spanned by computational basis states that are equivalent under the rewriting rules, to which the dynamics is constrained by construction. Combining the adiabatic time scaling with the first-order Trotterization error bound yields a total number of Trotter steps scaling as $N = \mathcal{O}(\Delta_{\min}^{-4})$. The resulting circuit depth can thus be bounded by
\begin{equation}
D = \mathcal{O}\left(\Delta_{\min}^{-4} \left(2w \cdot n_r + L\right)\right)\,.
\end{equation}
This scaling can be easily improved through a variety of techniques, including higher-order Trotterization schemes, counterdiabatic driving and other shortcuts to adiabaticity, as well as heuristics better suited to near-term quantum hardware, such as the Quantum Approximate Optimization Algorithm.

\subsection*{Fidelity between orbit states on quantum annealers}

Here, we show how the fidelity between orbit states can be measured on quantum annealers.
\\We prepare the orbit states $\ket{X_{S',\omega'}}$ and $\ket{X_{S'',\omega''}}$ on a quantum annealer. To this aim, we initialize the annealer respectively in the states $\ket{\omega'}$ and $\ket{\omega''}$. We evolve these initial states with the Hamiltonians $H_{S',\omega'}(t)$ and $H_{S'',\omega''}(t)$ for large enough times $\tau'$ and $\tau''$. In the adiabatic regime, we obtain
\begin{equation}
      \ket{X_{S',\omega'}} = \mathcal{T}\left[\mathrm{e}^{-i\int_0^{\tau'} dt H_{S',\omega'}(t)}\right]\ket{\omega'}\;,
\end{equation}
and
\begin{equation}
      \ket{X_{S'',\omega''}} = \mathcal{T}\left[\mathrm{e}^{-i\int_0^{\tau''} dt H_{S'',\omega''}(t)}\right]\ket{\omega''}\;,
\end{equation}
where $\mathcal{T}$ is the time-ordering operator.

Considering that $\mathcal{T}\left[\mathrm{e}^{-i\int_0^{\tau'} dt H_{S',\omega'}(t)}\right]^\dag = \mathcal{T}\left[\mathrm{e}^{-i\int_0^{\tau'} dt H_{S',\omega'}(\tau'-t)}\right]$, the fidelity between the orbit states is
\begin{align}\label{eq:analog_fidelity}
    |\langle X_{S'',\omega''}\ket{X_{S',\omega'}}|^2&=\Big|\bra{\omega'}\mathcal{T}\left[\mathrm{e}^{-i\int_0^{\tau'} dt H_{S',\omega'}(\tau'-t)}\right]\cdot\nonumber\\
    &\cdot\mathcal{T}\left[\mathrm{e}^{-i\int_0^{\tau''} dt H_{S'',\omega''}(t)}\right]\ket{\omega''}\Big|^2\;.
\end{align}

Now, we define the state
\begin{align}
    \ket{\omega''_{S',S''}}&=\mathcal{T}\left[\mathrm{e}^{-i\int_0^{\tau'} dt H_{S',\omega'}(\tau'-t)}\right]\nonumber\\
    &\cdot\mathcal{T}\left[\mathrm{e}^{-i\int_0^{\tau''} dt H_{S'',\omega''}(t)}\right]\ket{\omega''}\;.
\end{align}
This state can be prepared on the quantum annealer by evolving the initial state $\ket{\omega''}$ first with Hamiltonian $H_{S'',\omega''}(t)$ for a time $\tau''$, and then with the Hamiltonian $H_{S',\omega'}(\tau'-t)$ for a time $\tau'$.

Once $\ket{\omega''_{S',S''}}$ has been prepared, the fidelity in Eq.~(\ref{eq:analog_fidelity}) is measured as the expectation value of the projector $P_{\omega'} = \ket{\omega'}\bra{\omega'}$ onto the computational basis state $\ket{\omega'}$, i.e.,
\begin{equation}
    |\langle X_{S'',\omega''} \ket{X_{S',\omega'}}|^2 = \bra{\omega''_{S',S''}} P_{\omega'} \ket{\omega''_{S',S''}} \;.
\end{equation}
This quantity is the probability of sampling $\omega'$. It can be estimated by performing $N_s$ shots of the experiment and counting the frequency of outcomes corresponding to the configuration $\omega'$. The error on this estimation scales as $\mathcal{O}(1/\sqrt{N_s})$.

\subsection*{Comparison to classical approaches}

Here, we analyze connections and differences between quantum normal form reduction and the main classical approaches to the word problem and the counting problem. This comparison clarifies in which limits quantum normal form reduction can be regarded as a quantum extension of classical methods, and how insights from these approaches may be leveraged to improve quantum normal form reduction.

\textbf{Graph exploration}

The most direct classical approach is based on explicit graph-exploration algorithms, such as breadth-first search with memoization~\cite{10.5555/1614191}. These algorithms sequentially explore the connected component of the rewriting graph containing the input string, allowing one to test equivalence and to enumerate or count all connected words. Since the size of the connected subgraph typically grows exponentially with the size of the rewriting system or the input, such approaches quickly become computationally infeasible.

Adiabatically preparing the ground state of the Laplacian associated with the rewriting system corresponds to exploring the graph of connected words, but in quantum superposition rather than via sequential traversal. From this perspective, quantum normal form reduction is more naturally compared to random walks on graphs, whose equilibrium distribution is uniform over connected components. In both the classical random-walk and the quantum settings, the computational complexity is governed by the bottlenecks of the graph, which determine the spectral gap of the Laplacian. Overall, the possibility of a quadratic quantum speedup relative to classical random walks has been extensively studied in the literature, although its realization depends sensitively on structural properties of the specific rewriting system~\cite{Montanaro_2015}.

\textbf{Canonical form reduction}

More sophisticated approaches to the word problem are based on canonical form reduction~\cite{baader1998term}, that is, on constructing a procedure that maps all elements of the same equivalence class to a unique representative. This allows one to solve the word problem by comparing the normal forms associated with the input words. Quantum normal form reduction can be viewed as a quantum analogue of this paradigm, since it maps all equivalent strings to a single coherent quantum superposition. Unlike classical canonical forms, this superposition can address the counting problem, since it encodes information about the entire equivalence class.

Completion-based methods, most notably the Knuth--Bendix algorithm~\cite{Knuth1983}, provide a systematic way to obtain such canonical forms by transforming the original rewriting system into one that is both terminating and confluent. Completion can fail or not terminate, limiting the applicability of this approach. In contrast to quantum normal form reduction, the Knuth--Bendix procedure requires the user to specify a suitable reduction ordering, and its success is highly sensitive to this choice. In Section \textit{Knuth-Bendix algorithm} of this Supplementary Text, we compare the performance of the Knuth–Bendix algorithm with our approach for a specific case.

\textbf{Automata-based methods}

Automata-based methods consist of constructing an accepting automaton, i.e., a finite-state automaton that accepts exactly the strings connected from a given input, thereby solving the word problem~\cite{10.5555/1196416}. Since each accepting path through the automaton corresponds to a distinct string, dynamic programming techniques can be used to count connected words.

The tensor-network representation of orbit states provides a bridge between quantum normal form reduction and automata-based constructions. Indeed, matrix product states are equivalent to weighted finite-state automata that compute functions on strings, where the bond dimension controls the number of internal automaton configurations. Thus, orbit states can be interpreted as accepting automata with exponentially large expressive power, since a generic quantum state corresponds to a matrix product state with a bond dimension exponential in the system size. Quantum normal form reduction defines a systematic procedure for implicitly constructing such automata.

\textbf{Reduction to SAT}

Boolean satisfiability (SAT)–based methods typically encode the predicate ``$\omega_1$ rewrites to $\omega_2$ within $k$ rewrite steps'' as a Boolean formula that is satisfiable if and only if such a derivation exists~\cite{10.1007/978-3-642-11266-9_63}. This encoding enables the use of highly optimized SAT solvers to address instances of the word problem. Furthermore, reductions to SAT allow one to enumerate satisfying assignments using blocking clauses, although counting an exponentially large set of connected strings in this way generally requires an exponential number of solver calls. Approximate counting can be achieved with a polynomial number of SAT solver invocations using hashing-based techniques. These methods estimate the number of satisfying assignments of a Boolean formula by randomly partitioning the solution space in buckets using XOR constraints, and then exactly counting the solutions in a randomly selected bucket whose expected size is bounded by a fixed threshold~\cite{approxMC}.

Since the Laplacian can be written as a frustration-free sum of positive semi-definite operators $r^2 - r$, preparing orbit states is a quantum analogue of SAT-based approaches, but with non-commuting clauses~\cite{localHamComplexity}. While the annealing-based scheme proposed here should therefore be regarded as a heuristic that may be effective in typical instances, exploring connections with modern SAT solver heuristics could improve this approach.

Unlike the solution of a SAT encoding, the ground state of the Laplacian is unique and naturally encodes a coherent superposition of all classical solutions. Estimating the number of connected strings nevertheless still requires repeated runs of the algorithm to sample the overlap between the orbit state and the uniform superposition over connected strings~\cite{PhysRevLett.87.167902}. An analogue of approximate counting would be to restrict the Hilbert space to randomly chosen buckets, implemented by adding suitable penalty terms to the Laplacian operator.

\subsection*{Knuth-Bendix algorithm}

The Knuth-Bendix algorithm~\cite{Knuth1983} is a state-of-the-art classical approach for solving the word problem. This procedure transforms the original rewriting system $S$ into a new and non-invertible system $S_C$, whose rules increase a specified total order on strings, such as lexicographic order. The new system is equivalent to the original one, meaning that two strings are connected by $S$ if and only if they are connected by $S_C$. Moreover, the transformed system $S_C$ is both \emph{terminating}, meaning that no infinite sequence of rule applications is possible, and \emph{confluent}, meaning that any sequence of valid rule applications yields the same final result, regardless of the order in which rules are applied.

For a rewriting system that is both confluent and terminating, two strings $\omega'$ and $\omega''$ are equivalent if and only if repeated application of rewriting rules — regardless of the order in which they are applied — reduces both to the same string. This unique representative is called the \emph{normal form}. Reduction to normal form thus enables efficient resolution of the word problem, provided that such a system $S_C$ can be constructed.

The Knuth-Bendix algorithm is not guaranteed to terminate: in general, the construction of $S_C$ may fail, reflecting the undecidability of the word problem for arbitrary rewriting systems. However, a fair comparison with the Knuth-Bendix algorithm for the purposes of this work must account for the finite size of the strings. This constraint limits the maximum length of the rewriting rules generated during the execution of the algorithm, ensuring termination within finite time and memory resources that depend on $L$.

\begin{figure}[t!]
\includegraphics[width=0.5\linewidth]{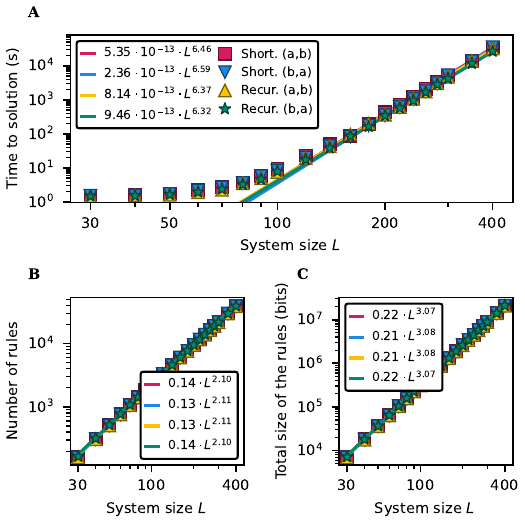}
\caption{
\textbf{Computational cost of the Knuth-Bendix algorithm as a function of the string size $L$ for different orderings.}
\textit{Panel~A:} Computational time required to construct a confluent string rewriting system, along with the corresponding polynomial fit for $L \geq 150$.
\textit{Panel~B:} Number of rules generated in the confluent string rewriting system, together with the corresponding polynomial fit.
\textit{Panel~C:} Total size of the rules in the confluent string rewriting system, along with the corresponding polynomial fit.}
\label{fig:knuth_bendix}
\end{figure}

As a benchmark for our quantum algorithm, we use the computer algebra system \textit{GAP}~\cite{GAP4} to run the Knuth-Bendix algorithm for the string rewriting system in Eq.~(\ref{eq:srs_def}). The algorithm is executed for both the \textit{shortlex} and \textit{recursive} orderings, and for both possible permutations of the alphabet, $(a,b)$ and $(b,a)$ (see GAP documentation for further details). All executions were performed on a virtual machine equipped with 6 Intel(R) Core(TM) i5-8500 CPUs and 16 GB of memory. The total execution time until termination, as well as the memory footprint of the resulting confluent rewriting system $S_C$, are reported in Figure~\ref{fig:knuth_bendix} for string lengths up to $L = 400$. Across different choices of ordering, the computational time scales asymptotically as $\mathcal{O}(L^{\sim 6.4})$. The number of rules in $S_C$ grows as $\mathcal{O}(L^{\sim 2.1})$, while the total memory required to store the system — measured by the cumulative length of all rules — scales as $\mathcal{O}(L^{\sim 3.0})$.

\subsection*{Computational complexity of the imaginary quantum annealing}

Here, we bound the computational complexity of imaginary quantum annealing (IQA) with respect to the energy gap and the fidelity susceptibility of the driving Hamiltonian ground state.

As in the main text, we consider IQA discretized in $N$ steps. We fix the time duration $\delta_\tau$ of each step, so that the total annealing time is $\tau = \delta_\tau N$. 

The system's Hamiltonian at the step $s$ is
\begin{equation}
    \hat{H}_s = \left[s\delta\hat{\mathcal{L}}_S - \left(1-s\delta\right) \ket{\tilde{\omega}}\bra{\tilde{\omega}}\right]\;,
\end{equation}
where $\delta = 1/N$ is the variation of the Hamiltonian parameter per step.

We call $\ket{\psi_s}$ the state of the system at step $s\in[0,\dots,N]$, $\ket{0_s}$ the ground state of the Hamiltonian $\hat{H}_s$, and $E_{n,s}$ the $n$-th energy level of $\hat{H}_s$.

We measure the error in approximating the ground state at a step $s$ as the infidelity $1-F_s$ between $\ket{\psi_s}$ and $\ket{0_s}$, where the fidelity $F_s$ is defined as
\begin{equation}
    F_s = \left|\langle 0_s\ket{\psi_s}\right|^2\;.
\end{equation}

The state evolution at each step is
\begin{equation}
    \ket{\psi_{s+1}} = \frac{\mathrm{e}^{-H_{s+1}\delta_\tau}\ket{\psi_{s}}}{\big|\big|\mathrm{e}^{-H_{s+1}\delta_\tau}\ket{\psi_{s}}\big|\big|}\;,
\end{equation}
so that the fidelity is
\begin{equation}\label{supp_eq:fidelity_def}
    \left|\langle 0_{s+1}\ket{\psi_{s+1}}\right|^2 = \frac{\left|\bra{0_{s+1}}\mathrm{e}^{-H_{s+1}\delta_\tau}\ket{\psi_{s}}\right|^2}{\bra{\psi_{s}}\mathrm{e}^{-2H_{s+1}\delta_\tau}\ket{\psi_{s}}}\;.
\end{equation}

The numerator of the last equation can be written as
\begin{equation}\label{supp_eq:fidelity_num}
    \big|\bra{0_{s+1}}\mathrm{e}^{-H_{s+1}\delta_\tau}\ket{\psi_{s}} \big|^2= \mathrm{e}^{-2E_{0,s+1}\delta_\tau}\big|\bra{0_{s+1}}\psi_{s}\rangle\big|^2\;.
\end{equation}

Let $P_s = \ket{0_s}\bra{0_s}$ be the projector on the ground state $\ket{0_{s}}$, and $P_s^\perp = 1-\ket{0_s}\bra{0_s}$ its orthogonal complement. We have:
\begin{equation}
    \mathrm{e}^{-H_{s+1}\delta_\tau}\ket{\psi_{s}} = \mathrm{e}^{-H_{s+1}\delta_\tau}\left(P_{s+1}+P_{s+1}^\perp\right)\ket{\psi_{s}}\;,
\end{equation}
so that the denominator becomes
\begin{align}\label{supp_eq:fidelity_den}
    \big|\big|\mathrm{e}^{-H_{s+1}\delta_\tau}\ket{\psi_{s}}\big|\big|^2 &= \bra{\psi_{s}} \left(P_{s+1}+P_{s+1}^\perp\right)\mathrm{e}^{-2H_{s+1}\delta_\tau}\left(P_{s+1}+P_{s+1}^\perp\right)\ket{\psi_{s}}\nonumber\\
    &= \bra{\psi_{s}} P_{s+1}\mathrm{e}^{-2H_{s+1}\delta_\tau}P_{s+1}\ket{\psi_{s}} + \bra{\psi_{s}}P_{s+1}^\perp \mathrm{e}^{-2H_{s+1}\delta_\tau}P_{s+1}^\perp\ket{\psi_{s}}\nonumber\\
    & \leq \mathrm{e}^{-2E_{0,s+1}\delta_\tau}\bra{\psi_{s}} P_{s+1}\ket{\psi_{s}} + \mathrm{e}^{-2E_{1,s+1}\delta_\tau}\bra{\psi_{s}} P_{s+1}^\perp\ket{\psi_{s}}\nonumber\\
    & = \mathrm{e}^{-2E_{0,s+1}\delta_\tau}\big|\langle 0_{s+1}\ket{\psi_{s}}\big|^2 + \mathrm{e}^{-2E_{1,s+1}\delta_\tau}\left(1-\big|\langle 0_{s+1}\ket{\psi_{s}}\big|^2\right)\;,
\end{align}
where at the second line we exploited the equation  $P_{s+1}H_{s+1}P_{s+1}^\perp = 0$, and, at the third line, we consider that $\bra{\psi_{s}}P_{s+1} \mathrm{e}^{-2H_{s+1}\delta_\tau}P_{s+1}\ket{\psi_{s}} =  \bra{\psi_{s}}P_{s+1} \ket{\psi_{s}}\mathrm{e}^{-2E_{0,s+1}\delta_\tau}$ and  $\bra{\psi_{s}}P_{s+1}^\perp \mathrm{e}^{-2H_{s+1}\delta_\tau}P_{s+1}^\perp\ket{\psi_{s}}\leq \bra{\psi_{s}}P_{s+1}^\perp \ket{\psi_{s}}||P_{s+1}^\perp \mathrm{e}^{-2H_{s+1}\delta_\tau}P_{s+1}^\perp||_\text{op}\leq \bra{\psi_{s}}P_{s+1}^\perp \ket{\psi_{s}}\mathrm{e}^{-2E_{1,s+1}\delta_\tau}$, where $||A||_\text{op}$ is the operator norm of $A$.

Substituting the corresponding terms in Eq.~(\ref{supp_eq:fidelity_def}) with the results of Eq.~(\ref{supp_eq:fidelity_num}) and Eq.~(\ref{supp_eq:fidelity_den}), we obtain
\begin{align}
    F_{s+1}&\geq \frac{\mathrm{e}^{-2E_{0,s+1}\delta_\tau}\big|\bra{0_{s+1}}\psi_{s}\rangle\big|^2}{\mathrm{e}^{-2E_{0,s+1}\delta_\tau}\big|\langle 0_{s+1}\ket{\psi_{s}}\big|^2 + \mathrm{e}^{-2E_{1,s+1}\delta_\tau}\left(1-\big|\langle 0_{s+1}\ket{\psi_{s}}\big|^2\right)}\nonumber\\
    &=\frac{\big|\bra{0_{s+1}}\psi_{s}\rangle\big|^2}{\big|\bra{0_{s+1}}\psi_{s}\rangle\big|^2 + \mathrm{e}^{-2\Delta_{s+1}\delta_\tau}\left(1-\big|\langle 0_{s+1}\ket{\psi_{s}}\big|^2\right)}\;,
\end{align}
where $\Delta_{s+1}$ is the first energy gap of $\hat H_{s+1}$.

Thus, the infidelity can be bounded as
\begin{align}\label{suppe_eq:partial}
    1 - F_{s+1} &\leq \frac{\mathrm{e}^{-2\Delta_{s+1}\delta_\tau}\left(1-\big|\langle 0_{s+1}\ket{\psi_{s}}\big|^2\right)}{\big|\bra{0_{s+1}}\psi_{s}\rangle\big|^2 + \mathrm{e}^{-2\Delta_{s+1}\delta_\tau}\left(1-\big|\langle 0_{s+1}\ket{\psi_{s}}\big|^2\right)}\nonumber\\
    &\leq \mathrm{e}^{-2\Delta_{s+1}\delta_\tau}\frac{1-\big|\langle 0_{s+1}\ket{\psi_{s}}\big|^2}{\big|\langle 0_{s+1}\ket{\psi_{s}}\big|^2}\;.
\end{align}

Now, we want to relate the infidelity at the step $s+1$ to the infidelity at the step $s$.\\
First, we decompose $\ket{\psi_s}$ on its parallel and orthogonal component with respect to $\ket{0_s}$, that is, $\ket{\psi_s} = \sqrt{F_s}\ket{0_s} + \mathrm{e}^{i\theta_s}\sqrt{1-F_s}\ket{e_s}$ for some complex phase $\theta_s$. Thus we have
\begin{align}\label{supp_eq:expand_0}
    \bra{0_{s+1}}\psi_{s}\rangle &= \sqrt{F_s}\bra{0_{s+1}}0_s\rangle +  \mathrm{e}^{i\theta_s}\sqrt{1-F_s}\bra{0_{s+1}}e_s\rangle\;.
\end{align}

At this point, we introduce the fidelity susceptibility $f_s$ for the ground state path of $H_s$ as a measure of the infinitesimal variation of the ground state~\cite{PhysRevLett.99.095701}, i.e.:
\begin{equation}
    |\bra{0_{s+1}}0_s\rangle|^2 = 1 - f_s \delta^2\;.
\end{equation}
which implies
\begin{equation}\label{supp_eq:expand_1}
    \ket{0_{s+1}} = \sqrt{1 - f_s \delta^2}\ket{0_s} + \sqrt{f_s}\delta \mathrm{e}^{i\theta'_s}\ket{\varepsilon_s}\;,
\end{equation}
where $\ket{\varepsilon}$ is the orthogonal part of $\ket{0_{s+1}}$ with respect to $\ket{0_s}$ and $\theta'_s$ is a complex phase.

Equation~(\ref{supp_eq:expand_1}) holds whenever the ground state path $\ket{0_s}$ is derivable. Being the Hamiltonian $H_s$ derivable, this happens whenever there is no level crossing between the first and the second energy level. In the case under exam here, where $H_s$ has negative non-diagonal entries, the Perron–Frobenius theorem guarantees the uniqueness of the ground state at each $s$ and therefore the absence of level crossing. Considering together  Eq.~(\ref{supp_eq:expand_0})  and Eq.~(\ref{supp_eq:expand_1}) we obtain
\begin{equation}
    \left|\bra{0_{s+1}}\psi_{s}\rangle\right|^2 = \left|\sqrt{F_s}\sqrt{1 - f_s \delta^2} +  \mathrm{e}^{i\theta_s}\mathrm{e}^{i\theta'_s}\sqrt{1-F_s}\sqrt{f_s}\delta \bra{e_s}e_s\rangle\right|^2\;.
\end{equation}

Exploiting the triangular inequality and considering that for small enough $\delta$, that is for large $N$, the modulus of the second addend is smaller than the modulus of the first, we obtain
\begin{equation}\label{supp_eq:overlap}
    \left|\bra{0_{s+1}}\psi_{s}\rangle\right|^2 \geq \left|\sqrt{F_s}\sqrt{1 - f_s \delta^2} - \sqrt{1-F_s}\sqrt{f_s}\delta |\bra{e_s}e_s\rangle|\right|^2\geq F_s(1 - f_s \delta^2)\;.
\end{equation}

Now we substitute Eq.~(\ref{supp_eq:overlap}) in Eq.~(\ref{suppe_eq:partial}) to obtain the relationship between fidelity at successive steps for large $N$:
\begin{align}
    1 - F_{s+1} \leq \mathrm{e}^{-2\Delta_{s+1}\delta_\tau}\frac{1-F_s(1 - f_s \delta^2)}{F_s(1 - f_s \delta^2)}\approx \mathrm{e}^{-2\Delta_{s+1}\delta_\tau}\left(\frac{1-F_s}{F_s} + \frac{f_s}{F_s^2}\delta^2\right)
\end{align}

Finally, we iteratively apply the last equation to obtain the evolution of the infidelity.\\
Considering that at the step $s=0$ the system is in the exact ground state of the Hamiltonian, so that $F_0=1$, we have
\begin{align}
    1 - F_{1} &\leq \mathrm{e}^{-2\Delta_{1}\delta_\tau}f_0\delta^2\;.
\end{align}
and for $s=2$, considering small $\delta$, we have
\begin{align}
    1 - F_{2} &\leq \mathrm{e}^{-2\Delta_{2}\delta_\tau}\left(\frac{\mathrm{e}^{-2\Delta_{1}\delta_\tau}f_0\delta^2}{1-\mathrm{e}^{-2\Delta_{1}\delta_\tau}f_0\delta^2} + \frac{f_1}{\left(1-\mathrm{e}^{-2\Delta_{1}\delta_\tau}f_0\delta^2\right)^2}\delta^2\right)\nonumber\\
    &\approx \mathrm{e}^{-2\Delta_{2}\delta_\tau}\left(\mathrm{e}^{-2\Delta_{1}\delta_\tau}f_0\delta^2 + f_1\delta^2\right) \nonumber\\
    &= \mathrm{e}^{-2\delta_\tau \left(\Delta_1+\Delta_2\right)}f_0\delta^2 + \mathrm{e}^{-2\delta_\tau \left(\Delta_2\right)}f_1\delta^2
\end{align}
Iterating this process, we get the following upper-bound for the final fidelity:
\begin{align}\label{supp_eq:main}
    1 - F_{N} \leq \sum_{s=1}^{N} \mathrm{e}^{-2\delta_\tau\sum_{i=s}^N\Delta_{i}}f_{s-1}\delta^2 = \frac{1}{N^2}\sum_{s=1}^{N} \mathrm{e}^{-2\delta_\tau\sum_{i=s}^{N}\Delta_{i}}f_{s-1}\;.
\end{align}

Equation~(\ref{supp_eq:main}) can be interpreted as follows. At each time step, an error proportional to the fidelity susceptibility of the ground state path is accumulated. This error is exponentially damped down during all the remaining evolution at a rate proportional to the instantaneous first energy gap.

Now, we recall that by construction, the operator $\hat H_s$ can be written in block diagonal form, where each block acts on a connected subgraph, i.e., on a single equivalence class. The dynamic is thus restricted to the subspace spanned by the equivalence class of the input word $\tilde\omega$. Thus, given the input word $\tilde\omega$, the fidelity susceptibility refers to the path of ground states in the corresponding sector of the Hilbert space, as well as the first energy gap.

\subsection*{Computational complexity and final energy gap}

Equation~(\ref{supp_eq:main}) is a bound that depends on the specific path of ground states in the exam, and, ultimately, on the input state $\ket{\tilde{\omega}}$. A less strong bound can be derived that does not depend on $\tilde{\omega}$ but only on the first energy gap of the Laplacian in the block where $\tilde{\omega}$ belongs. In the following, we call this gap the final gap $\Delta_{\tilde\omega}$.

First of all, we observe that in the last part of the dynamics, the accumulated error is damped at a rate that only depends on the gap of the Hamiltonian for $s \approx N$. Since the Hamiltonian norm is bounded, this gap can not deviate too much from $\Delta_{\tilde\omega}$. This observation can be placed on a more rigorous mathematical footing using Weyl’s perturbation theorem~\cite{bhatia1997matrix}, which establishes that the rate of change of eigenvalues is bounded by the operator norm of the variation of the Hamiltonian as
\begin{equation}\label{supp_eq:lipsc_0}
    |E_{i,s}-E_{i,N}| \leq ||H_s-H_{N}||_\text{op} =\left|\left| \frac{N-s}{N}\left(\hat{\mathcal{L}}_S +  \ket{\tilde{\omega}}\bra{\tilde{\omega}}\right)\right|\right|_\text{op}=\frac{N-s}{N}\left|\left| \hat{\mathcal{L}}_S +  \ket{\tilde{\omega}}\bra{\tilde{\omega}}\right|\right|_\text{op}\;.
\end{equation}

We can upper-bound the Hamiltonian norm as
\begin{equation}\label{supp_eq:laplacian_bound}
    ||\hat{\mathcal{L}}_S + \ket{\tilde{\omega}}\bra{\tilde{\omega}}||_\text{op}\leq \sum_r||\hat r^2-r||_\text{op}+||\ket{\tilde{\omega}}\bra{\tilde{\omega}}||_\text{op}\leq 2n_r+1\;
\end{equation}
where we used the triangular inequality for the operator norm, $n_r$ is the number of operators $\hat r^2-\hat r$ in the Laplacian, and each operator $\hat r^2-\hat r$ has norm $2$. Thus, Eq.~(\ref{supp_eq:lipsc_0}) becomes
\begin{equation}\label{supp_eq:lipsc}
    |E_{i,s}-E_{i,N}| \leq  \frac{N-s}{N}(2n_r+1)\;,
\end{equation}
which implies
\begin{equation}
    \Delta_{s} > \Delta_{\tilde\omega} - \frac{N-s}{N}(4n_r+2)\;.
\end{equation}
Let $s_*$ be the minimum value of $s$ for which the last expression is positive, that is
\begin{equation}
    s_* = N - \frac{N\Delta_{\tilde\omega}}{4n_r+2}\;.
\end{equation}
For $s<s_*$ we have
\begin{equation}
    \mathrm{e}^{-2\delta_\tau\sum_{i=s}^{N}\Delta_{i}}\leq \mathrm{e}^{-2\delta_\tau\sum_{i=s_*}^{N}\Delta_{i}}\leq \mathrm{e}^{-2\delta_\tau\sum_{i=s_*}^{N}\left(\Delta_{\tilde\omega} - \frac{N-i}{N}(4n_r+2)\right)}=\mathrm{e}^{-N\frac{\delta_\tau\Delta_{\tilde\omega}^2}{4n_r+2}}\,.
\end{equation}
and for $s\geq s*$
\begin{equation}
    \mathrm{e}^{-2\delta_\tau\sum_{i=s}^{N}\Delta_{i}}\leq \mathrm{e}^{-2\delta_\tau\sum_{i=s}^{N}\left(\Delta_{\tilde\omega} - \frac{N-i}{N}(4n_r+2)\right)}=\mathrm{e}^{-2\delta_\tau\left( (N-s)\Delta_{\tilde\omega} - \frac{(N-s)^2}{2N}(4n_r+2)\right)}\,.
\end{equation}

We substitute these two bounds in Eq.~(\ref{supp_eq:main}) to obtain
\begin{align}
    1 - F_{N} &\leq \frac{1}{N^2}\left(\sum_{s=1}^{s_*-1} f_{s-1} \mathrm{e}^{-N\frac{\delta_\tau\Delta_{\tilde\omega}^2}{4n_r+2}}+\sum_{s=s_*}^{N}f_{s-1} \mathrm{e}^{-2\delta_\tau\left( (N-s)\Delta_{\tilde\omega} - \frac{(N-s)^2}{2N}(4n_r+2)\right)}\right)
\end{align}
For large $N$, the second summation can be written as an integral. With a change of variable, we obtain
\begin{align}
    1 - F_{N} &\leq \frac{f_\text{avg}}{N}\mathrm{e}^{-N\frac{\delta_\tau\Delta_{\tilde\omega}^2}{4n_r+2}}+\frac{f_f}{N}\int_0^1 \mathrm{e}^{-\frac{2\delta_\tau N \Delta_{\tilde\omega}}{4n_r+2}( x - x^2)}dx\;.
\end{align}
For large $N$, we can use Laplace's method for approximating the integral by expanding the exponent around the endpoints of the domain, thus obtaining:
\begin{align}\label{eq:mmmm}
    1 - F_{N} &\leq \frac{f_\text{avg}}{N}\mathrm{e}^{-\frac{N\delta_\tau\Delta_{\tilde\omega}^2}{4n_r+2}}+\frac{f_f}{N}\frac{4n_r+2}{N\delta_\tau\Delta}\,.
\end{align}

Now we bound the fidelity susceptibility at the end of the process with respect to the final energy gap~\cite{PhysRevLett.99.095701} as
\begin{equation}
    f_s = \sum_{n \neq 0} \frac{\left|\langle n_s|\partial_s H_s|0_s\rangle\right|^2}{(E_{n,s}-E_{0,s})^2}\leq \frac{||\partial_s H_s||_\text{op}}{\Delta_{\tilde\omega}^2},
\end{equation}
which, in our case, considering the bound on the Laplacian norm in Eq.~(\ref{supp_eq:laplacian_bound}), becomes
\begin{equation}
    f_s \leq \frac{4n_r+2}{\Delta_{\tilde\omega}^2}.
\end{equation}

Replacing the bound on the final fidelity in Eq.~(\ref{eq:mmmm}) we obtain
\begin{align}
    1 - F_{N} &\leq \frac{f_\text{avg}}{N}\mathrm{e}^{-\frac{N\delta_\tau\Delta_{\tilde\omega}^2}{4n_r+2}}+\frac{(4n_r+2)^2}{N^2\delta_\tau\Delta^3}\,.
\end{align}

Then, the number of time steps needed to get a final infidelity $1-F_N\leq\epsilon$ is
\begin{equation}
    N = \mathcal{O}\left(\frac{n_r}{\Delta_{\tilde\omega}^2} \log\left( \frac{f_\text{avg}}{\epsilon}\right)\right) + \mathcal{O}\left(\sqrt{\frac{n_r^2}{\epsilon \delta_\tau\Delta_{\tilde\omega}^3}}\right)\;.
\end{equation}

Thus, when the number of rules in the rewriting system grows polynomially, the computational complexity scales polynomially with the inverse of the final minimum energy gap. This result holds even if the fidelity susceptibility diverges exponentially during the evolution, as a consequence of an exponentially closing gap. By contrast, this advantage is lost in real-time quantum annealing: due to the unitary nature of the dynamics, errors accumulated at intermediate times cannot be dissipated at the end of the process.

\end{document}